\input phyzzx

\def\Real{{\bf R}}
\def\Zahl{{\bf Z}}
\def\d{{\partial}}
\def\Del{{\mit \Delta}}
\def\del{{ \delta}}
\def\vnab{{\vec \nabla}}
\def\Ord{{{\cal O}}}
\def\<{{\langle}}
\def\>{{\rangle}}
\def\equi{{\ \Longleftrightarrow \ }}
\def\det{{\rm det}}
\def\=by#1{{\mathrel{\mathop=^{#1}}}}
\def\thinlayer{{\ \mathrel{\mathop{\longrightarrow}^{\scriptscriptstyle \rm thin}_{\scriptscriptstyle \rm layer}}\ }}

\def\bP{{\bf P}}

\def\Hbar{{\overline{H}}}

\def\tG{{\widetilde{G}}}
\def\tB{{\widetilde{B}}}
\def\tK{{\widetilde{K}}}
\def\tH{{\widetilde{H}}}
\def\tY{{\widetilde{Y}}}
\def\tR{{\widetilde{R}}}
\def\tGam{{\widetilde{\Gamma}}}

\def\vB{{\vec B}}
\def\vH{{\vec H}}

\def\vN{{\vec N}}

\def\vn{{\vec n}}
\def\vx{{\vec x}}
\def\vy{{\vec y}}
\def\val{{\vec \alpha}}
\def\vA{{\vec A}}
\def\vX{{\vec X}}
\def\vF{{\vec F}}

\def\Gam{{{\Gamma}}}
\def\Lam{{\Lambda}}
\def\kap{{\kappa}}

\def\A{{\scriptscriptstyle  A}}
\def\B{{\scriptscriptstyle  B}}
\def\D{{\scriptscriptstyle  D}}
\def\E{{\scriptscriptstyle  E}}
\def\G{{\scriptscriptstyle  G}}
\def\H{{\scriptscriptstyle  H}}
\def\a{{ a}}
\def\b{{ b}}
\def\c{{ c}}
\def\rd{{ d}}
\def\e{{ e}}
\def\f{{ f}}
\def\i{{ i}}
\def\j{{ j}}
\def\k{{ k}}
\def\l{{ l}}
\def\m{{ m}}
\def\n{{ n}}
\def\U{{\scriptscriptstyle  U}}
\def\V{{\scriptscriptstyle  V}}
\def\W{{\scriptscriptstyle  W}}
\def\X{{\scriptscriptstyle  X}}
\def\Y{{\scriptscriptstyle  Y}}
\def\Z{{\scriptscriptstyle  Z}}

\def\XA{{X^\A}}
\def\xa{{x^\A}}
\def\qb{{q^\b}}

\def\qU{{q^\U}}
\def\tBbetA{{\tB_\beta^{\ \A}}}
\def\tBalA{{\tB_\alpha^{\ \A}}}
\def\BbA{{B_\b^{\ \A}}}

\def\Rp{{\Real^p}}
\def\Mn{{\bf M}^n}
\def\Mp{{\bf M}^p}
\def\M1{{\bf M}^1}
\def\Mtwo{{\bf M}^2}
\def\NMn{{\Xi(\Mn)}}
\def\Lag{{\cal L}}

\def\QMn{{Q_{\downarrow \Mn}}}

\def\Sph{{\bf S}}

\def\cA{{N}}
\def\F{{\cal F}}
\def\To{{ \buildrel \circ \over T  }}

\def\yo{{ \buildrel {\ \circ} \over y  }}
\def\to{{ \buildrel {\ \circ} \over t  }}
\def\etas#1{{\eta_{\kern -.1em \lower .8ex \hbox{$\scriptstyle #1$}}}}
\def\ssub#1{{s_{\kern -.1em \lower .8ex \hbox{$\scriptstyle #1$}}}}
\def\dsub#1{{\d_{\kern -.1em \lower .8ex \hbox{$\scriptstyle #1$}}}}
\def\nablAst{{ \buildrel * \over \nabla  }}

\def\Gamf{{ \buildrel {\scriptscriptstyle (f)} \over {\Gam^m} }}
\def\GamabeAst{{ \buildrel { * \quad}  \over {\Gam^e_{ab}} }}

\def\Heff{{H_{\rm  eff}}}
\def\Keff{{K_{\rm  eff}}}

\def\Tp{{T'}}

\def\Curv{{C_{\vB}}}
\def\SolA{{\Omega_{\vB}}}
\def\Surf{{S_{\vB}}}
\def\vAso{{\vA_{\rm sole}}}
\def\Gauss{{\rm \scriptscriptstyle (Gauss)}}

\def\PB{{\}_{\kern -.2em \lower .7ex\hbox{$\scriptscriptstyle {\rm PB}$}}}}


\REF\BJOR{
{J.~D.~Bjorken}\journal{Ann. of Phys.}&24(63){174};
{T.~Eguchi and H.~Sugawara}\journal{Phys. Rev. D}&10(74){4257};
{K.~Kikkawa}\journal{Prog. Theor. Phys.}&56(76){947}.
}

\REF\CREM{
{E.~Cremmer and  B.~Julia}\journal{Nucl. Phys.}&B159(79){141}.}

\REF\DADDA{
{A.~D'Adda, M.~L\"usher and P.~Divecchia}\journal{Nucl. Phys.}&B146(78){63};
\journal{}&B152(79){125};
E.~Witten\journal{Nucl. Phys.}&B149(79){285};
K.~B.~Bitar\journal{Phys. Rev.}&D24(81){2654};
B.~Milewski, {\it Four dimensional theories with composite gauge fields and their supersymmetric generalizations} (Warsaw Univ., 1982).
}

\REF\BANDO{
{M.~Bando, T.~Kugo, S.~Uehara, K.~Yamawaki and T.~Yanagida}\journal{Phys. Rev. Lett.}&54(85){1215};
{M.~Bando, T.~Kugo and  K.~Yamawaki}\journal{Nucl. Phys.}&B259(85){473};
 \journal{Prog. Theor. Phys.}&73(85){1541};
 \journal{Phys. Rep.}&164(88){217}.
}

\REF\FUJIA{
{K.~Fujii, N.~Ogawa, K.-I.~Sato, N.~Chepilko, A.~Kobushkin and T.~Okazaki}\journal{Phys. Rev. D}&44(91){3237}.
}

\REF\KIKKAWA{
{K.~Kikkawa}\journal{Phys. Lett.}&B297(92){89};
{K.~Kikkawa and H.~Tamura}\journal{Int. J. Mod. Phys.}&A10(95){1597};
{T.~Hatsuda and H.~Kuratsuji}, {\it Gauge Field Emerging from Extra Dimensions---a Born--Oppenheimer approach---}, University of Tsukuba Preprint UTHEP-286~(1994).
}

\REF\WEINB{
S.~Weinberg and E.~Witten\journal{Phys. Lett.}&96B(80){59};
T.~Kugo\journal{Phys. Lett.}&109B(82){205}.
}

\REF\TAKAGI{
{S.~Takagi and T.~Tanzawa}\journal{Prog. Theor. Phys.}&87(92){561}.
See also {Y.~Nagoshi and S.~Takagi}\journal{J. of Phys.}&A24(91){4093}.
}

\REF\FUJIB{
{K.~Fujii and N.~Ogawa}\journal{Prog. Theor. Phys.}&89(93){575};
K.~Fujii, in {\it Hadrons and Nuclei from QCD}, eds. K.~Fujii, Y.~Akashi and B.~L.~Reznik (World Scientific, Singapore, 1994).
}

\REF\MARA{
P.~Maraner and C.~Destri\journal{Mod. Phys. Lett.}&A8(93){891};
P.~Maraner\journal{J. Phys.}&A28(95){2939}.
}

\REF\MARAB{
P.~Maraner\journal{Ann. of Phys.}&246(96){325}.
}

\REF\ARNOLD{
For example, V.~I.~Arnol'd, {\it Mathematical Methods of Classical Mechanics} (Nauka, Moscow, 1974).
See also N.~G.~van Kampen and J.~J.~Lodder\journal{Am. J. Phys.}&52(84){419}.
}

\REF\BORN{
M.~Born and J.~R.~Oppenheimer\journal{Ann. der Physik}&84(27){457}.
See, e.g., A.~Messiah, {\it M\'ecanique Quantique} (Dunod, Paris, 1959).
}

\REF\BERRY{
{M.~V.~Berry}\journal{Proc. R. Soc. London}&A392(84){45}.
Relevant papers on the Berry phase are found in
{\it Geometrical Phases in Physics}, eds. {A.~Shapere and  F.~Wilczek} (World Scientific, Singapore, 1989).
}

\REF\KOBAYASHI{
{S.~Kobayashi}, {Differential Geometry of Connections and Gauge Theory} (Shokabo, Tokyo, 1989); 
{S.~Kobayashi}, {\it Differential Geometry of Curves and Surfaces} (Shokabo, Tokyo, 1977);
S.~Sasaki, {\it Differential Geometry ---surface theory---} (Iwanami Shyoten, Tokyo, Japan).)
}

\REF\McMULLAN{
{D.~McMullan and I.~Tsutsui}\journal{Ann. of Phys.}&237(95){269}; 
\journal{Phys. Lett.}&B320(94){287}. \hfill \break
See also {S.~Tanimura and I.~Tsutsui}\journal{Mod. Phys. Lett.}&A10(95){2607}.
}

\REF\MACKEY{
{G.~W.~Mackey}, {\it Induced Representations of Groups and Quantum Mechanics} (Benjamin, New York, 1969).
}

\REF\LEVAY{
{P.~L\'evay, D.~McMullan and I.~Tsutsui}\journal{J. Math. Phys.}&37(96){625}.
}

\REF\OHNUKI{
{Y.~Ohnuki and S.~Kitakado}\journal{Mod. Phys. Lett.}&A7(92){2477}; \journal{J. Math. Phys.}&34(93){2827}.
K.~Fujii, S.~Kitakado and Y.~Ohnuki\journal{Mod. Phys. Lett.}&A10(95){867}.
}

\REF\LANDS{
{N.~P.~Landsman and N.~Linden}\journal{Nucl. Phys.}&B365(91){121};
{N.~P.~Landsman}\journal{Rev. Math. Phys.}&4(92){503}.
}

\REF\SuperG{
For example, Y.~Fujii, {\it Introduction to Theory of Supergravity}[in Japanese] (McGraw--Hill Book Company, Tokyo, 1987).
}

\REF\CHEPILKO{
{N.~M.~Chepilko and K.~Fujii}\journal{Yad. Fiz.}&58(95){1137};
\journal{Phys. At. Nucl.}&58(95){1063}.
}

\REF\RASHEV{
For example, {P.~K.~Rashevskii}, {\it Riemannian Geometry and Tensor Analysis} (Nauka, Moscow, 1967);
{Y.~Asanaga}, {\it Introduction to Riemannian Geometry} (Kyoritsu Syuppan, Tokyo, 1978).
}

\REF\TOLAR{
{J.~Tolar}, {\it On a Quantum Mechanical d'Alembert Principle}, Lecture Notes in Physics {\bf 313} (Springer--Verlag, Berlin, Heidelberg, 1988).
}

\REF\SUGANO{
{R.~Sugano}\journal{Prog. Theor. Phys.}&46(71){297};
{T.~Kimura and R.~Sugano}\journal{Prog. Theor. Phys.}&47(93){1004}.
}

\REF\FUJIC{
{K.~Fujii, K.-I.~Sato, N.~Toyota and A.~Kobushkin}\journal{Phys. Rev. Lett.}&58(87){651};
{K.~Fujii, A.~Kobushkin, K.-I.~Sato and N.~Toyota}\journal{Phys. Rev. D}&35(87){1896}.
}

\REF\MIYAZ{
K.~Fujii and H.~Miyazaki\journal{Prog. Theor. Phys.}&93(95){823}.
}

\REF\TANIM{
S.~Tanimura\journal{Ann. of Phys.}&220(92){229}.
}

\REF\JENSEN{
{H.~Jensen and H.~Koppe}\journal{Ann. of Phys.}&63(71){586};
{R.~C.~T.~da~Costa}\journal{Phys. Rev. A}&23(81){1982};
P.~Exner, P.~Seba and P.~Stovicek\journal{Phys. Lett.}&A150(90){179};
N.~Ogawa\journal{Prog. Theor. Phys.}&87(92){513};
{M.~Ikegami and Y.~Nagaoka}\journal{Prog. Theor. Phys. Suppl.}&106(91){235};
C.~Destri, P.~Maraner and E.~Onofri\journal{Nuovo Cim.}&A107(94){237}.
}

\REF\KOBU{
H.~P.~Berg\journal{Lett. Math. Phys.}&6(82){183};
T.~Homma, T.~Inamoto and T.~Miyazaki\journal{Phys. Rev. D}&42(90){2049};
N.~Ogawa, K.~Fujii and A.~Kobushkin\journal{Prog. Theor. Phys.}&85(90){894};
{A.~Kobushkin, N.~Ogawa, K.~Fujii and N.~M.~Chepilko}\journal{Yad. Fiz.}&52(90){772};
{N.~Ogawa, K.~Fujii, N.~Chepilko and A.~Kobushkin}\journal{Prog. Theor. Phys.}&85(91){1189}.
}

\REF\FUJID{
The main contents in the first two subsections in Sec. 3 have been described in the paper, {K.~Fujii and S.~Uchiyama}\journal{Prog. Theor. Phys.}&95(96){461}.
}

\REF\KUGLER{
{M.~Kugler and S.~Shtrikman}\journal{Phys. Rev. D}&37(88){934}.
}

\REF\WEINBb{
For example, S.~Weinberg, {\it Gravitation and Cosmology --- Principles and applications of the general theory of relativity} (John Wily \& Sons, New York$\cdot$ \break
Chichester$\cdot$Brisbane$\cdot$Toronto, 1972).
}

\REF\YANG{
See, e.g., {C.~N.~Yang}, {\it Gauge Fields} in {\it Proceedings of the Hawaii Topical Conference in Particle  Physics 1975}, (University of Hawaii, Hawaii, 1976) ed. {P.~N.~Dobson};  
{L.~C.~Biedenharn and J.~D.~Louck}, {\it The Racah--Wigner Algebra in Quantum Theory}, Encyclopedia of Mathematics and its Applications {\bf 9}, ed. Gian-Carlo Rota, (Addison-Wesley, London$\cdot$Amsterdam$\cdot$Don Mills$\cdot$Ontario$\cdot$Sydney$\cdot$Tokyo, 1981).
}

\REF\HANNAY{
{J.~H.~Hannay}\journal{J. Phys.}&A18(85){221}.
}

\REF\CHIAO{
{R.~Y.~Chiao} and {Y.-S.~Wu}\journal{Phys. Rev. Lett.}&57(86){933};
{A.~Tomita  and  R.~Y.~Chiao}\journal{Phys. Rev. Lett.}&57(86){937}. 
}

\REF\WIGNER{
E.~P.~Wigner\journal{Ann. Math.}&40(39){149};
Y.~Ohnuki, {\it Unitary Representation of the Poincar\'e Group and Relativistic Wave Functions} (World Scientific, 1988).
}

\REF\OGAWA{
N.~Ogawa, {\it Quantum mechanical embedding of spinning particle and induced spin-connection}, Hokkaido University of Education Preprint, Jan. 1997.
}

\REF\CHRIST{{N.~H.~Christ and T.~D.~Lee}\journal{Phys. Rev. D}&12(75){1606};
{N.~H.~Christ}\journal{Physics Reports (Section C of Phys. Letters)}&23(76){297}. 
}

\REF\GERVAIS{ {J.-L.~Gervais  and  A.~Jevicki}\journal{Nucl. Phys.}&B110(76){93}.
See also the papers in \journal{Physics Reports (Section C of Phys. Letters)}&23(76){}. 
}

\REF\KUGO{{T.~Kugo}\journal{Soryushiron Kenkyu}&54(77){95}[in Japanese].
}

\REF\RAJA{
T.~Rajaraman, {\it Solitons and Instantons} (North-Holland, Amsterdam$\cdot$New York$\cdot$Oxford, 1982).
}

\REF\WEYL{{H.~Weyl}\journal{Z. Physik}&46(27){1};  \hfill\break
{M.~M.~Mizrahi}\journal{Jour. Math. Phys.}&16(75){2201}. \hfill\break
See, e.g., {B.~Sakita}, {\it Quantum Theory of Many-Variable Systems and Fields}, World Scientific Lecture Notes in Physics {\bf 1}, (World Scientific, Singapore, 1985).
}

\FIG\SETUP{
An illustration of the curvilinear coordinates $\{q^1, q^2, q^3\}$ in the case of $n=2$ and $p=3$.
}
\FIG\GaussMap{
The helix is mapped to a circle $C_{\vec B}$ under the Gauss' spherical map.
}


\pubnum{EPHOU-96-004(Revised)}

\titlepage

\title{Geometrically Induced Gauge Structure on Manifolds Embedded in a Higher Dimensional Space$\foot{\rm The main part of this paper was reported in XXI International Colloquium on Group Theoretical Methods in Physics, 15--20 July 1996, Goslar, Germany.}$}

\vskip 1.0cm

\centerline{\twelvecp Kanji  Fujii,$\foot{E-mail: kafujii@jpnyitp.yukawa.kyoto-u.ac.jp, fujii@particle.phys.hokudai.ac.jp}$ Naohisa Ogawa,$\foot{E-mail: ogawa@particle.phys.hokudai.ac.jp}$ Satoshi Uchiyama$\foot{E-mail: uchiyama@particle.phys.hokudai.ac.jp}$}

\address{ Department of Physics, Faculty of Science,\break
 Hokkaido University, Sapporo 060, Japan}

\vskip 0.5cm
\centerline{ and }
\vskip 0.5cm

\centerline{\twelvecp Nikolai Mikhailovich Chepilko$\foot{E-mail: chepiln@olinet.isf.kiev.ua}$}

\address{ Institute of Physics, Academy of Sciences of Ukraine, \break
 Nauki Prospect 46, Kiev 252650, Ukraine}

\vskip 1.0cm

\abstract{
We explain in a context different from that of Maraner the formalism for describing motion of a particle, under the influence of a confining potential, in a neighbourhood of an $n$-dimensional curved manifold ${\bf M}^n$ embedded in a $p$-dimensional Euclidean space ${\bf R}^p$ with $p\geq n+2$.
The effective Hamiltonian on ${\bf M}^n$ has a (generally non-Abelian) gauge structure determined by geometry of ${\bf M}^n$.
Such a gauge term is defined in terms of the vectors normal to ${\bf M}^n$, and its connection is called the ${N}$-connection.
This connection is nothing else but the connection induced from the normal connection of the submanifold $\Mn$ of $\Real^p$.
In order to see the global effect of this type of connections, the case of ${\bf M}^1$ embedded in $\Real^3$ is examined, where the relation of an integral of the gauge potential of the ${N}$-connection (i.e., the torsion) along a path in ${\bf M}^1$ to the Berry's phase is given through {Gauss}  mapping of the vector tangent to ${\bf M}^1$.
Through the same mapping in the case of ${\bf M}^1$ embedded in $\Real^p$, where the normal and the tangent quantities are exchanged, the relation of the ${N}$-connection to the induced gauge potential (the canonical connection of the second kind) on the $(p-1)$-dimensional sphere $\Sph^{p-1}$ ($p\geq 3$) found by Ohnuki and Kitakado is concretely established;
the former is the pull-back of the latter by the {Gauss} mapping.
Further, this latter which has the monopole-like structure is also proved  to be gauge-equivalent to the spin-connection of $\Sph^{p-1}$.
Thus the $\cA$-connection is also shown to coincide with the pull-back of the spin-connection of $\Sph^{p-1}$.
Finally, by extending formally the fundamental equations for ${\bf M}^n$ to infinite dimensional case,  the present formalism is applied to the field theory that admits  a soliton solution.
The resultant expression is in some respects different from that of Gervais and Jevicki.
}


\chapter{Introduction}

\noindent
The local gauge symmetry, regarded usually as the most fundamental physical principle, is incorporated from the outset in the starting Lagrangian, and is broken  through some mechanism such as Higgs mechanism; the effective Lagrangian at a low-energy region becomes less symmetric, in which some gauge bosons become massive.
Contrary to this mechanism, however, there exists  another logical possibility, in which some gauge symmetries are generated dynamically at a low-energy region {\it a posteriori}, and/or appear in the low-energy effective Lagrangian as a hidden local symmetry,$^{\BJOR - \KIKKAWA}$ although the problem how to generate the kinetic terms for dynamical gauge fields is yet in controversy.$^{\WEINB}$

In the present paper, we investigate in detail the induced gauge structure found in the confining-potential approach (shortly, the {\it confining approach})  to the constrained system on the $n$-dimensional curved manifold $\Mn$ embedded in a higher dimensional Euclidean space $\Real^p$.
Such an induced gauge structure, which depends on geometry of $\Mn$, is generated from the degrees of freedom in the normal directions to $\Mn$, on which a particle motion is confined by a steep confining potential in $\Real^p$ with $p\geq n+2$.
Although general feature of this induced gauge structure has been explained in essence in Refs.~${\FUJIB}$ and ${\MARA}$, it seems meaningful to discuss fully its basic formalism and to investigate its logical structures as well as extension to field theories in a context somewhat different from Ref.~${\MARAB}$;
we expect that such an investigation  may give a new clue to exploring further the second possibility mentioned in the preceding paragraph.

Here we give some remarks on the  geometrically induced gauge structure in the confining approach by including its short process of development.
In classical mechanics, a mechanical system constrained on $\Mn$ in $\Real^p$ with $(p-n)$ holonomic constraints can be equivalently defined as a system in $\Real^p$ with a very steep potential (with $(p-n)$-dimensions in the normal directions to $\Mn$) together with the $2(p-n)$ initial conditions.$^{\ARNOLD}$
As to the quantum version of the confining-potential approach to d'Alembert's principle,$^{\ARNOLD}$ various authors have devoted themselves for these 25 years
 to the study of forms of the $\hbar^2$-order contributions, called the {\it quantum potential}.
 (We will give its explicit form in the present approach in Sec.~2 together with related references.)
 
In the year 1992, Takagi and Tanzawa$^{\TAKAGI}$ have considered the particle motion in a 3-dimensional Euclidean space $\Real^3$, where a particle in motion is assumed to be confined in a thin tube along a curve $\M1$ under the influence of some confining potential.
They showed that in the effective Hamiltonian on $\M1$ there appears a certain kind of $U(1)$-gauge potential which depends on the torsion of $\M1$ and gives rise to Aharonov--Bohm-like effect (called the geometry-induced AB-like effect).
Two of the present authors (K.~F. and N.~O.),$^{\FUJIB}$ and independently Maraner and Destri$^{\MARA}$ have generalized the result obtained in Ref.~$\TAKAGI$ to the particle motion in a $p$-dimensional Euclidean space $\Rp$, where a particle in motion is confined in a thin layer along an $n$-dimensional manifold $\Mn$ (with $n+2 \leq p$) due to a confining potential.
Similarly to the case $\M1\subset \Real^3$, there appears, in the effective Hamiltonian on $\Mn$, a new connection, depending on the geometry of $\Mn$ and behaving as the gauge potential which is generally non-Abelian.
The terminology {\it thin layer} (in the case of $\M1$, a thin tube) means that geometric properties along the manifold $\Mn$ change slowly in comparison with those along directions normal to $\Mn$.
This condition corresponds logically to the Born--Oppenheimer approximation$^{\BORN}$ (or the adiabatic one).
Indeed, Maraner has pointed out$^{\MARAB}$ that the embedding formalism or the confining approach $^{\FUJIB, \MARA}$ to problems with holonomic constraints$^{\ARNOLD}$ can well describe quantum systems such as polyatomic molecules treated in the rigid body approximation, and then leads to Abelian or non-Abelian monopole gauge structure as well as the Berry phase$^{\BERRY}$ which is usually derived under the Born--Oppenheimer approximation.$^{\BORN}$

In the representation with respect to the quantum algebra with $(p-n)$ components wave functions, the geometrically induced gauge potential, which is denoted by $T_{\V \W, a}$ in this paper, is  the {\it normal fundamental form} in Ref. ${\MARAB}$.
Since $(p-n)$ components wave functions are identified with sections of the normal bundle ${T(\Mn)}^\bot$, this connection is what is called  the Van der Waerden--Bortolotti connection or the normal connection of $\Mn$ for the embedding into $\Real^p$ by mathematicians.$^{\KOBAYASHI}$
As we will see in Sec.~2, since $T_{\V \W, a}=-T_{\W \V, a}$, this normal connection induces a connection in the bundle of normal frames $O^\bot(\Mn)$ in a natural manner.
Such an induced connection leads to a connection in a vector bundle associated with the principal bundle $O^\bot(\Mn)$.
For convenience, we call the obtained  connection  the {\it N-connection}.
Then we can say that the geometrically induced gauge potential is the $N$-connection.
In the first part of this paper, after presenting the formulation of the confining approach, we examine the induced gauge potential or the $\cA$-connection in the case of $\M1\subset \Real^3$ for the purpose of seeing the situation how the nontrivial phase effect appears.
The {Gauss} mapping$^{\KOBAYASHI}$ of the vector $\vB_1$, tangent to $\M1$, is shown to be useful.
The case of a helical $\M1$ is examined in detail, which may give us a possible way of experimental test of the $\cA$-connection.

Meanwhile, the important contribution to the quantum mechanics on a coset space $G/H$ has been done by McMullan and Tsutsui$^{\McMULLAN}$ on the basis of reformulation of the Mackey's quantization scheme$^{\MACKEY}$ on the coset space $G/H$, and they pointed out$^{\McMULLAN}$ that emergence of the gauge structure as well as the spin  are a natural consequence of the theory.
Further, L\'evay, McMullan and Tsutsui$^{\LEVAY}$ showed that the induced  gauge field found by Ohnuki and Kitakado$^{\OHNUKI}$ in the problem of $n$-dimensional sphere $\Sph^n$ embedded in $\Real^{n+1}$ is none other than the $H$-connection,$^{\LEVAY}$ observed by Landsman and Linden.$^{\LANDS}$
It was mentioned in Ref.~${\LEVAY}$, although the confining-potential approach$^{\TAKAGI -  \MARAB}$
 is quite different from Mackey's approach based on $G/H$, the geometrically induced gauge field found in Refs. ${\TAKAGI - \MARA}$ also appears to be of the type of the $H$-connection.

L\'evay et al.,$^{\LEVAY}$ however, have not given any concrete explanation on the second point.
In the confining approach, in order to obtain the induced gauge potential on a sphere $\Sph^n$ embedded in $\Real^p$, it is necessary to take $p\geq n+2$, whereas the induced gauge potential on $\Sph^n$ embedded in $\Real^{n+1}$ exists.$^{\OHNUKI}$
It seems worthwhile to make clear the relation between these two kinds of the gauge potentials.
The key to solve this problem lies in the {Gauss} mapping$^{\KOBAYASHI}$ of the vector $\vB_1$, tangent to $\M1$ embedded in $\Real^p$ with $p\geq 3$;
through this mapping the tangent and the normal quantities are exchanged.
In the following we show concretely the $\cA$-connection leads to the induced gauge field found by Ohnuki and Kitakado.$^{\OHNUKI}$
Further, this latter is also proved  to be gauge-equivalent to the spin-connection$^{\SuperG}$ of $\Sph^{p-1}$;
the spin-connection on $\Sph^{p-1}$ ($p\geq 3$) is seen to have the monopole-like gauge structure.

The aims of the present paper are summarized as follows:
(i) To present in an elementary way the formulation of the confining approach,$^{\FUJIB, \MARA}$ (ii) in a context somewhat different from that in Maraner's papers,$^{\MARA, \MARAB}$, to examine the relation of the $\cA$-connection to the monopole-like structure and the Berry phase, specifically in the case of $\M1\subset \Real^3$, (iii) to establish in the case of $\M1 \subset \Real^p$ (with $p\geq 3$) the relation of the $\cA$-connection to the induced gauge field on $\Sph^{p-1}$ found in Ref.~${\OHNUKI}$, and finally (iv) to extend the formalism of the confining approach to the scalar field theory that allows soliton solutions.

The following considerations are organized as follows.
In Sec.~2, we summarize the formulation of the confining approach in Ref.~$\FUJIB$.
In Sec.~3, detailed structures of the $\cA$-connection are to be examined specifically in the case of $\M1\subset \Real^3$.
We give a simple example which represents the relation of the $\cA$-connection to the monopole field as well as the Berry phase.
In Sec.~4, the relation of the $\cA$-connection in the case of $\M1\subset \Real^p$ (with $p\geq 3$) to the induced gauge field on $\Sph^{p-1}$ found by Ohnuki and Kitakado$^{\OHNUKI}$ is examined through the {Gauss} mapping.
In Sec.~5, the formulation given in Sec.~2 is extended, as its possible application, to the scalar field theory allowing soliton solutions.
Finally, Sec.~6 is devoted to additional remarks and discussions.

\chapter{Particle Motion in a Neighbourhood of $\Mn$ Embedded in $\Rp$}
\section{Fundamental equations of a hypersurface}

\noindent
In this section, we summarize the results in the preceding paper,$^{\FUJIB, \CHEPILKO}$ so that the present paper becomes selfcontained.

$\Real^p$ is regarded as a $p$-dimensional Euclidean space in a natural manner, and let $\Mn$ be an $n$-dimensional smooth manifold embedded in $\Real^p$ as shown in Fig.~1.

\centerline{$\lower 6ex \hbox to 150pt {\hrulefill}$}
\centerline{$\hbox to 150pt {\hrulefill}$}
\centerline{Fig.~1}
\centerline{$\hbox to 150pt {\hrulefill}$}
\centerline{$\lower -6ex \hbox to 150pt {\hrulefill}$}

\noindent
We denote the natural coordinate system on $\Real^p$ by $\{ X^\A \ ; \ A=1, \ldots, p \}$, and we introduce a coordinate system on a neighbourhood of $\Mn$ in the following way.
Let (${\bf U}$, $\{ q^b \ ; \ b=1, \ldots, n \}$) be a local coordinate system on $\Mn$.
At each point $q\in {\bf U}\subset \Mn$, there is a linearly independent set $\{\vN_\U= (N_\U^\A (q)) \ ; \ U=n+1, \ldots , p \}$ of $p$-dimensional orthogonal unit vectors that are normal to $\Mn$ at a point $q$.
Take a sufficiently small and appropriate neighbourhood $\Xi({\bf U})$ of ${\bf U}$ in $\Real^p$ such that the following condition is satisfied:
 for each point $Q\in \Xi({\bf U})$ there is a unique point $q\in {\bf U}\subset\Mn$, which is denoted by $\QMn$, such that $\vX(Q)-\vX(q)=(X^\A(Q)-X^\A(q))$ is normal to $\Mn$ at $q=\QMn$.
$\vX(Q)-\vX(q)$ is expressed as a linear combination of $\vN_\U(q)$'s with the expansion coefficients $q^\U$ ($U=n+1, \ldots, p$).
The set of $\{ q^b,\ q^\U \ ; \ b=1, \ldots, n, \ U=n+1, \ldots, p \}$ is thought to constitute a curvilinear coordinate system  on the neighbourhood $\Xi({\bf U})$ of ${\bf U}\subset\Mn$.
Each member of $\{ q^{n+1}, \ldots, q^p\}$ is called {\it the normal coordinate} in the frame of the normal vectors $\{ \vN_{n+1}, \ldots, \vN_{p} \}$.
Put $\NMn=\bigcup \{ \Xi({\bf U})$ $|$ {\bf U} is a coordinate neighbourhood of $\Mn$ $\}$ $\subset \Real^p$.
Then each set $\{ q^b,\ q^\U \ ; \ b=1, \ldots, n, \ U=n+1, \ldots, p \}$ on $\Xi({\bf U})$ is a local coordinate system of the  $p$-dimensional submanifold $\NMn$ of  $\Real^p$.
By putting $x^\A(q^b):=X^\A(q)$ with $q=\QMn$, a coordinate transformation  from $\{ q^b, \ q^\U \}$ into $\{ X^\A \}$ is given as
$$
\XA(\qb, \qU)=\xa(\qb)+\sum_{\U=n+1}^p \qU N_\U^\A(\qb), \quad A=1, \ldots, p.
\eqn\twoOne
$$
Hereafter we utilize the notation of indices as follows:
\item{1)} small Latin indices $\a$, $\b$, $\ldots$ run from 1 to $n$, and are used to represent components of curvilinear coordinates on $\Mn$;
\item{2)} Greek indices run from 1 to $p$, and are used to unify two sets $\{ q^b \}$ and $\{ \qU\}$ by writing the union as $\{ q^\beta \}$;
\item{3)} the final part of capital letters $U$, $V$, $\ldots$ run from $n+1$ to $p$;
\item{4)} the first part of capital letters $A$, $B$, $\ldots$ run from 1 to $p$, and represent the vector property in the Euclidean space $\Rp$.

\noindent
For simplicity, we omit  symbols of summation, say $\sum_{\U=n+1}^p$ for a dummy index $U$;
we identify each point in $\NMn$ with its local curvilinear-coordinates  $\{ q^\beta \}$ and write $Q=(q^\beta)=(q^b,~ q^\U)$, $q=(q^b)=\QMn$  and so on.

We define vector fields $\tB_\beta$ ($\beta=1, \ldots, p$) and $\vB_b$ ($b=1, \ldots, n$) on $\NMn$ as
$$
\tBbetA(Q):={\d \XA(Q) \over \d q^\beta}, \qquad
\BbA(q):={\d x^\A(q) \over \d q^b}, \quad q=\QMn  \eqn\twoTwoA
$$
The vectors  $\vB_\b(q):=(\BbA(q))$ ($\b$=1, $\ldots$ $n$) in $\Rp$ are tangent to $\Mn$ at a point $(x^\A(q))\in\Mn\subset \NMn$.
Components of $\tB_\beta(Q)$ are explicitly written as 
$$
 \tB_b^{\ \A}(Q)=\BbA(q)+q^\W\d_\b N_\W^\A(q),\quad \tB_\U^{\ \A}(Q) =N_\U^\A(q).
\eqn\twoTwoB
$$

The metric tensor on $\NMn\subset\Rp$ for the curvilinear coordinates $\{ q^\beta \}$, which we write as 
$\tG_{\alpha \beta}(Q)$, is given by
$$
\tG_{\alpha \beta}(Q)=\tBalA(Q) \etas{\A\D}\tB_\beta^{\ \D}(Q), \quad (\etas{\A\D})  = \hbox{diag} ( +1, \ +1, \ \cdots, \ +1 ).  \eqn\twoThreeA
$$
The metric tensor on $\Mn$ for the curvilinear coordinates $\{ q^b \}$, which we write as $g_{\a\b}$, is given by 
$$
g_{\a \b}(q)=B_\a^{\ \A}(q) \etas{\A\D} B_\b^{\ \D}(q).  \eqn\twoThreeB
$$

The fundamental equations for $\BbA$ and $N_\V^{\ \A}$ are$^{\RASHEV}$
$$
\d_\a\BbA = \Gam^\rd_{\a\b}B_\rd^{\ \A}+H_{\W\a\b} \eta^{\W\U}N_\U^{\ \A}, 
\eqn\twoFourA
$$
$$
\d_\a N_\V^{\ \A} = -H_{\V\a}^{\ \ \rd}B_\rd^{\ \A}-T_{\V\W, \a} \eta^{\W\U}N_\U^{\ \A}, 
\eqn\twoFourB
$$
where $\Gam^\rd_{\a\b}$ is Christoffel symbol constructed in terms of $g_{a b}$;
$H_{\W\a\b}(q)=H_{\W\b\a}(q)$
($=H_{\W\b}^{\ \ \ d}(q)$ $g_{d\a}(q)$ );
$T_{\V\W, \a}(q)=-T_{\W\V, \a}(q).$
Equations $\twoFourA$ and $\twoFourB$ are the extended forms of generalized Frenet--Serret equations in the case of $\M1\subset \Real^p$ to the present case of $\Mn\subset \Rp$.(See Appendix B.)

\section{Structure of the metric $\tG_{\alpha \beta}$}

\noindent
By substituting $\twoTwoB$ into $\twoThreeA$ and taking account of $\twoFourB$,  concrete form of $\tG_{\alpha \beta}$ is given as follows:
$$
\left( \tG_{\alpha \beta}\right)
=\left(\matrix{
\tG_{\a \b} & \tG_{\a \U} \cr
\tG_{\V \b} & \tG_{\V \U}
}\right)     \eqn\twoFiveA
$$
with 
$$ \eqalign{
\tG_{\a \b}&=\lambda_{\a\b}+\tG_{\a \X}\eta^{\X\Y}\tG_{\b\Y}, \cr
\lambda_{\a \b}:&=g_{\a\b}-2H_{\W\a\b}\,q^\W+q^\X q^\Y H_{\X\a}^{\ \ \ \e}\, H_{\Y\e\b}, \cr
\tG_{\a \U}&=\tG_{\U \a}=T_{\U\X, \a}\, q^\X, \quad \tG_{\V\U}=\etas{\V\U}.
}
$$
Since $(\tG_{\alpha \beta})$ is symmetric and $\tG_{\V \U} = \etas{\V \U}$, the inverse $(\tG^{\alpha \beta})$ is given as follows:
$$
\left( \tG^{\alpha \beta}\right)
=\left(\matrix{
\tG^{\a \b} & \tG^{\a \U} \cr
\tG^{\V \b} & \tG^{\V \U}
}\right)     \eqn\twoFiveB
$$
with 
$$ \eqalign{
\tG^{\a \b}&=\lambda^{\a\b}, \quad \lambda^{\a\b}\lambda_{\b\rd}=\delta^\a_{\ \rd}, \cr
\tG^{\a \U}&=\tG^{\U \a}=-\lambda^{\a\rd}T_{\X\W, \rd}\, q^\W \eta^{\X\U},\cr
\tG^{\V \U}&=\eta^{\V\U}+\eta^{\V\X}T_{\X\W, \b}\, q^\W \lambda^{\b\rd}T_{\Y\Z,\rd}\, q^\Z\eta^{\Y\U}.
}
$$
The above expressions of $\tG_{\alpha \beta}$ and $\tG^{\alpha \beta}$ are obtained with no approximation.$^{\MARA, \CHEPILKO}$
If we take the thin-layer approximation,$^{\FUJIB}$ i.e.,
$$
\epsilon_H:=\sum_{a, b}\left| H_{\U\a\b}\,  q^\U\right| \ll 1 \hbox{ and } \epsilon_T:=\sum_{b, \U}\left| T_{\U\V, \b}\, q^\V\right| \ll 1, \eqn\twoSixA
$$
we obtain
$$
\lambda^{\b\rd}=g^{\b\rd}+2H_\V^{\ \b\rd}q^\V+3H_{\V\a}^{\ \ \b}H_\W^{\ \a\rd}q^\V q^\W+\Ord[(\epsilon_H+\epsilon_T)^3].
\eqn\twoSixB
$$

The integrability condition $\d_\a \d_\b B_c^{\ \A}=\d_\b \d_\a B_c^{\ \A}$ leads to the curvature tensor on $\Mn$ expressed in terms of $H_{\V\a\b}$;
$$\eqalign{
R_{\a\b, \c\rd}  :&= g_{\a\e}( \d_\c\Gam^{\e}_{\ \rd\b}-\d_\rd\Gam^{\e}_{\ \c\b}+\Gam^{\e}_{\ \c\f} \Gam^{\f}_{\ \rd\b}-\Gam^{\e}_{\ \rd\f} \Gam^{\f}_{\ \c\b})   \cr
&= \etas{\A\B} (H^\A_{\ \a\c} H^\B_{\ \b\rd}-H^\A_{\ \a\rd} H^\B_{\ \b\c})
}\eqn\twoSevenA
$$
with $H^\A_{\ \a\b}:=H_{\V\a\b}\, \eta^{\V\W}N_\W^{\ \A}$ (Euler--Schouten tensor$^{\RASHEV}$);
hence, Ricci tensor and the scalar curvature are respectively given by
$$
R_{\b\c}=g^{\a\rd}R_{\a\b, \c\rd}=\etas{\A\B}(H^{\A \ \rd}_{\ \b} H^\B_{\ \rd\c}-H^{\A \ \rd}_{\ \rd} H^\B_{\ \b\c}),
\eqn\twoSevenB
$$
$$
R=g^{\b\c} R_{\b\c}=\etas{\A\B} (H^{\A \ \rd}_{\ \b} H^{\B \ \b}_{\ \rd}-H^{\A \ \b}_{\ \b} H^{\B \ \rd}_{\ \rd}).
\eqn\twoSevenC
$$
(The sign of $R$ for a sphere becomes negative in the present definition.)

Similarly, the integrability condition $\d_\a \d_\b N_\V^{\ \A}-\d_\b \d_\a N_\V^{\ \A}=0$ leads to
$$\eqalign{
-&\d_\rd H_{\V\a}^{\ \ \b}+\d_\a H_{\V\rd}^{\ \ \b}-H_{\V\a}^{\ \ \c}\Gam^\b_{\rd\c}+H_{\V\rd}^{\ \ \c}\Gam^\b_{\a\c} \cr
&\quad +H_{\X\rd}^{\ \ \ \b}\eta^{\X\W}T_{\V\W, \a}-H_{\X\a}^{\ \ \ \b}\eta^{\X\W}T_{\V\W, \rd}=0,
} \eqn\twoEightA
$$
$$\eqalign{
R_{\rd\a, \V\W}:=& -\d_\rd T_{\V\W, \a}+\d_\a T_{\V\W, \rd}+T_{\X\V, \rd}\, \eta^{\X\Y}T_{\Y\W, \a}-T_{\X\V, \a}\, \eta^{\X\Y}T_{\Y\W, \rd}  \cr
=& -H_{\V\rd}^{\ \ \ \b} H_{\W\b\a}+H_{\V\a}^{\ \ \ \b} H_{\W\b\rd}.
}  \eqn\twoEightB
$$

We define the {\it extrinsic mean curvature} $H$ $^{\RASHEV, \TOLAR}$  by
$$
H:={1\over n}\left[ H_{\U\b}^{\ \ \ \b}\eta^{\U\V}H_{\V\rd}^{\ \ \ \rd}\right]^{1/2},  \eqn\twoNineA
$$
then the scalar curvature $R$, $\twoSevenC$, is rewritten as
$$
R=H^{\A \ \rd}_{\ \b} \etas{\A\B} H^{\B \ \b}_{\ \rd}-n^2 H^2. \eqn\twoNineB
$$
The meaning of $H$ is seen when we notice that $H$ is equal to the {\it curvature} appearing in Frenet--Serret equation in the case of $\M1\subset \Real^3$ and that $H$ is equal to the magnitude of the mean curvature vector $\left(\Hbar^\A\right)$ defined by
$$
\Hbar^\A:={1\over n}\sum_{\f=1}^n H^\A_{\ \ \b\rd}\, v_{(\f)}^\b\, v_{(\f)}^\rd. 
\eqn\twoTenA
$$
Here, $\{ \sum_{d}v_{(1)}^\rd(q)(\d/\d q^d)_q, \ldots, \sum_{d}v_{(n)}^\rd(q)(\d/\d q^d)_q \}$ is a set of unit tangent vectors at a point $q$ in $\Mn$ that are orthogonal to each other, i.e.,
$$
g_{\b\rd}\, v_{(\f)}^\b \, v_{(\e)}^\rd=\etas{\f\e}.
\eqn\twoTenB
$$
In fact, since $\sum_{f}v_{(f)}^b v_{(f)}^d=g^{b d}$, we have 
$$
 \Hbar^\A = {1\over n} H^\A_{\ \ \b\rd} g^{\b\rd}={1\over n} H^{\A \ b}_{\ b};
$$
consequently,
$$
 \left[ \Hbar^\A \etas{\A\B} \Hbar^\B \right]^{1/2}=H.
 \eqn\twoTenC
$$

\section{Canonical quantization}

\noindent
Let us consider a particle (with its mass = 1) in $\Rp$, the motion of which is confined in a neighbourhood of $\Mn$ in the presence of some potential $V(X^\A)$.
By denoting its kinetic energy by $\tK$, the Lagrangian is given as
$$
\Lag=\tK-V(X^\A).     \eqn\twoLagrange
$$
In order to investigate the quantum-mechanical structure of the kinetic energy
$$
\tK:={1\over 2}\dot X^\A \etas{\A\B} \dot X^\B \qquad \left( \hbox{with } \dot X^\A={d X^\A\over d \tau} \right),      \eqn\twoEleven
$$
expressed in terms of the $q^\beta$-variables, we adopt the quantization procedure developed in Refs. ${\SUGANO}$, ${\FUJIC}$, which is consistent if the curvilinear coordinates  $\{ q^\beta \}$ can be transformed into  the Euclidean coordinates (hence the curvature $\tR$ vanishes) and the metric tensor is $\dot q^\beta$-independent as in the present case.
By employing this procedure which enables us to treat coordinate transformations algebraically, we can perform all calculations in the operator formalism quantum-mechanically from  the outset without recourse to the differential operator representation of momenta, and obtain the kinetic term $\twoEleven$ expressed in terms of operators of curvilinear coordinates and their conjugate momenta.
In the following, we denote a symmetrized product of quantum-mechanical dynamical variables $A$ and $B$ by $\< A, B \>$, i.e.,
$$
\< A, B \>:={1\over 2}(A B + B A).
$$

We assume 
$$
[ q^\beta, \dot q^\delta ]=i\hbar f^{\beta\delta}(q^\gamma), \quad [ q^\beta, q^\delta ]=0,  \eqn\twoTwelveA
$$
where $f^{\beta \delta}$ is a function of only $q^\gamma$'s  and not of $\dot q^\gamma$'s.
The time derivative of $X^\A(q^\gamma)$ is defined as$^{\SUGANO, \FUJIC}$
$$
\dot X^\A(q^\gamma):=\< \tB_\alpha^\A(q^\gamma), \dot q^\alpha \>.
\eqn\twoTwelveB
$$
This definition ensures  the hermiticity and the quantum-mechanical covariance.$^{\MIYAZ}$ 
The momentum operator $P_\A$, conjugate to $X^\A$, is given by
$$
P_\A={\d \Lag \over \d \dot X^\A}=\< \etas{\A\B}\, \tB_\beta^{\ \B},\ \dot q^\beta\>.     \eqn\aOne
$$
As explained shortly in Appendix A, we see
$$
f^{\alpha \beta}(q^\gamma)=\tG^{\alpha \beta}(q^\gamma)
\equi [ X^\A, P_\B]=i\hbar \delta^\A_\B.       \eqn\twoTwelveC
$$
The momentum operator $p_\beta$, conjugate to $q^\beta$, is written with use of (A.3) in Appendix A as
$$
p_\beta:={\d \Lag \over \d \dot q^\beta}=\< \etas{\A\B}\tB_\beta^\A, \< \tB_\alpha^\B, \dot q^\alpha \> \>=\< \tG_{\beta \alpha}, \dot q^\alpha\>.
\eqn\twoTwelveD
$$
From $\twoTwelveD$ we obtain due to the first relation of $\twoTwelveC$
$$
[q^\alpha, p_\beta]=i\hbar \delta^\alpha_\beta.
 \eqn\twoTwelveE
$$
$p_\alpha$ is reexpressed as
$$
p_\alpha  = \< \tB_\alpha^{\ \A},\  P_\A \>.   \eqn\twoTwelveDB
$$
From the explanation in Appendix A, we see
$$
[P_\A, P_\B]=0 \equi [p_\alpha, p_\beta]=0 \equi \F^{\alpha \beta}=0
\eqn\twoThirteenA
$$
under the integrability  condition
$$
\d_\beta \d_\alpha X^\A-\d_\alpha \d_\beta X^\B =0.   \eqn\twoThirteenB
$$
Here $\F^{\alpha \beta}$ is defined by Eq.~(A.6).$^{\MIYAZ, \TANIM}$
(Note that in Ref.~${\TANIM}$ the covariance under the general coordinate transformation is not fully taken into account.)
Thus we saw that under the assumptions $\twoTwelveB$ and $\twoTwelveA$ with the choice $f^{\alpha \beta}(q^\gamma)=\tG^{\alpha \beta}(q^\gamma)$, the canonical commutation  relations among $\{ q^\beta, \ p_\beta ; \ \beta=1, \ldots, p \}$ are equivalent to those among $\{ X^\A, P_\A; \ A=1, \ldots, p \}$.

\section{Structure of kinetic energy operator}
\noindent
It is easy to see the kinetic energy $\tK$ , Eq.~$\twoEleven$, is rewritten quantum-mechanically in the covariant form as 
$$\eqalign{
\tK &={1\over 2}p_\alpha \tG^{\alpha \beta} p_\beta -{1\over 2} \hbar^2 \tY(q^\gamma) \cr
&= {1\over 2}\tG^{-1/4} p_\alpha \tG^{1/2} \tG^{\alpha \beta}p_\beta \tG^{-1/4},}  \eqn\twoFourteenA
$$
where
$$
\tY(q^\gamma):=-{1\over 2}\d_\alpha (\tG^{\alpha \beta} \tGam_\beta)-{1\over 4}\tG^{\alpha \beta}\tGam_\alpha \tGam_\beta;
 \eqn\twoFourteenB
$$
$$\eqalign{
\tGam_\alpha :=& \tGam^\beta_{\ \alpha \beta}=(\d_\alpha \tG_{\rho \lambda})\tG^{\rho \lambda}/2=\d_\alpha \tG /(2\tG); \cr
\tG :=& \left| \det \tG_{\alpha \beta} \right|.
}
$$
By noting
$$
\tG=\left| \det(\lambda_{\a\b}) \det(\etas{\U\W}) \right|=\det(\lambda_{\a\b})=:\lambda,     \eqn\twoFifteenA
$$
and the structure of $(\tG^{\alpha \beta})$ given by $\twoFiveB$, we obtain
$$
\tK={1\over 2} \lambda^{-1/4}\pi_\a \lambda^{1/2} \lambda^{\a\b} \pi_\b \lambda^{-1/4}+{1\over 2} \lambda^{-1/4} p_\V \lambda^{1/2} \eta^{\V\W} p_\W \lambda^{-1/4}.    \eqn\twoFifteenB
$$
Here, $\pi_\a$ is defined as
$$
\pi_\a:=p_\a+ {1\over 2} T_{\V\W, \a} L^{\V\W}, \eqn\twoSixteenA
$$
$$
L^{\V\W}:=q^\V \eta^{\W\X}p_\X -q^\W\eta^{\V\X} p_\X.  \eqn\twoSixteenB
$$
The operator $L^{\V\W}$ satisfies the commutation relation
$$
[L^{\V\X}, L^{\W\Y}]=i\hbar(\eta^{\V\W}L^{\X\Y}+\eta^{\X\Y}L^{\V\W}-\eta^{\V\Y}L^{\X\W}-\eta^{\X\W}L^{\V\Y}).
\eqn\twoSixteenC
$$
In Sec.~6, in connection with the generalization to the case of $\Mn\subset \Mp$, we give a simple explanation why $\tK$ reduces to the form $\twoFifteenB$.

In the thin-layer approximation $\twoSixA$, we obtain
$$
\tK \thinlayer K^*=K+ {1\over 2}p_\V\eta^{\V\W} p_\W+\Del V^*,
\eqn\twoSeventeenA
$$
where 
$$
K:={1\over 2}g^{-1/4} \pi_\a g^{1/2} g^{\a\b} \pi_\b g^{-1/4}, \qquad g:=|\det(g_{\a\b})|,
\eqn\twoSeventeenB
$$
$$
\Del V^* ={\hbar^2\over 2}\left[-{1\over 2}H_{\U a}^{\ \ \ b}\eta^{\U\V}H_{\V b}^{\ \ \ a} +{1\over 4} n^2 H^2 \right]
={\hbar^2\over 2}\left[ -{R\over 2}-{1\over 4}n^2 H^2 \right];
\eqn\twoSeventeenC
$$
$H$ is the extrinsic mean curvature defined by $\twoNineA$.
Note that we have 
$$
\Del V^*=-{\hbar^2\over 2}\left( \tY(q, q^\U)-Y(q) \right) \vrule_{\lower 1ex\hbox{$\scriptscriptstyle \rm thin \ layer$}},  \eqn\twoEighteenA
$$
where $Y(q)$ is the quantity corresponding to $\twoFourteenB$ constructed in terms of $g_{\a\b}$ and rewritten as
$$
Y(q)=-{1\over 4}\left( R+ g^{\a\b}\Gam^\e_{\ \a\rd}\Gam^\rd_{\ \b\e} -\d_a \d_b g^{a b}\right);
\eqn\twoEighteenB
$$
$K$, $\twoSeventeenB$, can be expressed as
$$
K={1\over 2}\pi_\a g^{\a\b} \pi_\b -{\hbar^2\over 2}Y(q).
\eqn\twoEighteenC
$$

As to the so-called quantum potential $\Del V^*$, its form in the confining approach has been considered by many authors,$^{\TOLAR, \JENSEN}$ and compared with the corresponding form obtained in accordance with the Dirac approach to constrained dynamical systems.$^{\KOBU}$
An interesting view on $\Del V^*$ including the {\it extrinsic} quantity together with the {\it intrinsic} curvature is found in Maraner's paper.$^{\MARAB}$

The form of the effective kinetic term $\twoSeventeenB$ on $\Mn$ suggests existence of a certain gauge structure whose gauge potential is ${1\over 2}T_{\V \W, a} L^{\V \W}$ (see $\twoSixteenA$), caused by the `off-diagonal' elements $\tG_{\alpha \U}$ and $\tG^{\alpha \U}$ of the metric tensor $\tG_{\alpha \beta}$ and $\tG^{\alpha \beta}$.
We examine further details of such a gauge structure in the following subsection.

\section{Properties of $T_{\V\W, \a}$}

\noindent
Utilizing the commutation relation $\twoSixteenC$ and the definition of $R_{\a\b, \V\W}$ $\twoEightB$, we obtain$^{\CHEPILKO}$
$$
[\pi_\a, \pi_\b]={i\hbar\over 2}R_{\a\b, \V\W}L^{\V\W}.   \eqn\twoNineteenA
$$
This is analogous to the familiar case of a charged-particle motion in $\Real^3$ under the influence of magnetic field $\vH$, where we have
$$
[\pi_\j, \pi_\k]=i\hbar{e \over c}F_{\j\k};   \eqn\twoNineteenB
$$
$\pi_\j=p_\j-{e\over c}A_\j, \j$ = 1, 2, 3; $F_{\j\k}=\epsilon_{\j\k\l}H^\l$.
Also, in non-Abelian gauge theories, we have
$$
[\pi_\m, \pi_\n]=i\hbar{g \over 2}W_{\m\n,\j\k}G^{\j\k};   \eqn\twoNineteenC
$$
$$\eqalign{
\pi_\m &= p_\m+{g\over 2} A_{\j\k, \m}G^{\j\k}, \cr
W_{\m\n, \j\k} &= -\d_\m A_{\j\k, \n}+\d_\n A_{\j\k, \m} +g\left( A_{\i\j, \m}\, \eta^{\i \l}A_{\l\k, \n}-A_{\i\j, \n}\, \eta^{\i \l}A_{\l\k, \m} \right);
}
$$
$G^{\j\k}$'s are generators of the relevant gauge group that satisfy the same commutation relations as $\twoSixteenC$.
Thus we see that
$$
T_{\V\W, \a}=N_\V^{\ \A} \etas{\A\B} \d_\a N_\W^{\ \B}    \eqn\twoNineteenD
$$
has a complete analogy to the gauge potential, and its various aspects will be explored in the following  in the context different from that of Maraner.$^{\MARAB}$

We take the total Hamiltonian 
$$
\tH=\tK+V, \eqn\twoTwenty
$$
where $\tK$ is defined as $\twoFourteenA$, and $V$ includes a confining potential part responsible to confining the particle motion in the neighbourhood of $\Mn$, and assume $\tH$ is invariant under the rotation of a set of $\{ \vN_{n+1}, \ldots, \vN_p \}$ such as
$$
{N'}_\V^{\ \A}(q)=N_\W^{\ \A}(q) \Lam^{\W}_{\ \V}(q), \quad \Lam^{\X}_{\ \W}\, \etas{\X\Y} \Lam^{\Y}_{\ \V}=\etas{\W\V}.   \eqn\twoTwentyoneA
$$
Then, $T_{\W\V, \b}(q)$ is transformed under the $SO(p-n)$ rotation as
$$
T_{\W\V, \b}=N_\W^{\ \A}\etas{\A\B} \d_\b N_\V^{\ \B} \longmapsto 
\Tp_{\W\V, \b}=(\Lam^{-1})_\W^{\ \X} T_{\X\Y, \b} \Lam^\Y_{\ \V}+(\Lam^{-1})_{\W}^{\ \X}\etas{\X\Y} \d_\b \Lam^\Y_{\ \V}.     \eqn\twoTwentyoneB
$$
(For clarity, here we wrote the inverse matrix of $\Lam=(\Lam^\X_{\ \ \Y})$ as $\Lam^{-1}=((\Lam^{-1})_\Y^{\ \ \X})$.)
Thus, we can say that the  quantity $T_{\W \V, b}$ is a gauge potential and the $N$-connection emerges.
In Ref.~$\FUJIB$, we have illustrated through a simple example how the gauge structure appears in the effective Hamiltonian in Schr\"odinger equation for the particle motion in $\Mn$, where some degeneracy of states in the $q^\U$-space is assumed in order to obtain a set of suitable representations of $\{ T_{\V\W, \b}L^{\V \W} \}$.
More general case, where the full $SO(p-n)$ symmetry does not hold, are considered by Maraner.$^{\MARA, \MARAB}$

Although $T_{\V\W, \b}$ cannot be eliminated generally over the whole range of $\Mn$, it is important to examine how a nontrivial global effect, called the geometry-induced Aharonov--Bohm effect by Takagi and Tanzawa,$^{\TAKAGI}$ is brought about. \break
Maraner$^{\MARAB}$ has shown that, applying the present formalism to polyatomic mole-\break
cules with approximately rigid configurations, Abelian and non-Abelian mono- \break
pole-like structures are found.
Here it should be noted that, in the $\M1$ case, the {\it field strength} $R_{a b, \V \W}$ appearing in $\twoNineteenA$ vanishes identically even when $T_{\V\W, a}$ is not equal to zero,
while the nontrivial global contribution due to the $\cA$-connection, i.e., $T_{\V\W, a}$, is expected to exist, as noted by Takagi and Tanzawa.$^{\TAKAGI}$
In the following section we examine details of the gauge structure of $\M1$ case to see how nontrivial phase effects appear.

\chapter{Global Aspects of Geometrically Induced Gauge Structure}

\section{The Case of $\M1\subset \Rp$ with $p\geq 3$}

\noindent
As will be explained in the following, $T_{\W\V, \b}$ is eliminated locally;
thus it is necessary for us to examine the global structure brought about by $T_{\W\V, \b}$ in order to see physical effects due to the geometrically induced gauge.$^{\FUJID}$
In this connection, Takagi and Tanzawa$^\TAKAGI$ pointed out, in the case of a thin tube embedded in $\Real^3$, that a nonvanishing AB type effect arises when

$$
\int_0^l \tau(q)\, dq \not= 0.      \eqn\threeOne
$$
Here the parameter $q$ specifies a point on a center curve $\M1$ of the thin tube (i.e., $q=q^1$ in accordance with our present use of symbols);
$l$ is the length of this closed curve, where $0\leq q \leq l$;
$\tau$ is the torsion in Frenet--Serret equations of $\M1\subset \Real^3$ (See Eq.~(3.8).).
Remembering, however, the remark given at the end of the last section, it seems necessary for us to investigate the logical structure of $\threeOne$.
Indeed, we will see in the following two subsections that the condition for the curve $\M1$ to be closed in the real space  is not always necessary in order to obtain a geometrically induced phase such as $\threeOne$.
With the aim of exploring the situation further, we first examine the general case $\M1\subset \Rp$ with $p\geq 3$ as follows.

The fundamental equations for $\M1$ embedded in $\Rp$ are, from $\twoFourA$ and $\twoFourB$, written as
$$
{d \over d q^1} \vB_1 = \sum_{\W=2}^p h_\W \vN_\W,        \eqn\threeTwoA 
$$
$$
 {d \over d q^1}\vN_\V = -h_\V \vB_1 +\sum_{\W=2}^p \vN_\W T_{\W\V, 1},   \eqn\threeTwoB
$$
where $h_\W$:=$H_{\W 1}^{\ \ \ 1}$
;
the arrow represents the vector property in $\Rp$.
In the present and the following two subsections, the coordinate $q^1$ is simply written as $q$ which is taken to be the length on the curve $\M1$, that is, $\vB_1$ is a unit vector tangent to $\M1$.
$\threeTwoA$ and $\threeTwoB$ are expressed in the form, similar to Frenet--Serret equations (B.11) given in Appendix B, as
$$
{d \over d q}(\vB_1 \   \vN_2 \ \cdots \ \vN_p )=(\vB_1 \ \vN_2  \ \cdots \ \vN_p)\left(\matrix{ 0 & -h_2 & -h_3 & \cdots & -h_p \cr
   h_2 &    &   &        &   \cr
   h_3 &           & T &     &   \cr
   \vdots   &    &   &        &   \cr
   h_p &           &   &        &   \cr
   }\right)    \eqn\threeThreeA
$$
with 
$$
T:=\pmatrix{  &  &  \cr
              & T_{\V\W, 1} & \cr
              &  &  \cr},    \qquad  (V, W  = 2, \ldots, p).
               \eqn\threeThreeB
$$
Here, at each point $q\in \M1$ a set of normal vectors $\{ \vN_2, \ldots, \vN_p \}$ is chosen so that it is related with another set $\{ \vn_2, \ldots ,$ $ \vn_p \}$ defined in Appendix B by an orthogonal transformation $\Lam(q) \in SO(p-1)$;
$$
(\vN_2 \ \cdots \  \vN_p)=(\vn_2 \ \cdots \ \vn_p) \Lam  \qquad \hbox{ with   } \Lam^T=\Lam^{-1}.   \eqn\threeFour
$$
Comparing $\threeThreeA$ with (B.11), we obtain
$$
h_\U = \kap_1 \Lambda^2_{\ \U}, \qquad   U = 2, \ldots, p,     \eqn\threeFiveA 
$$
$$
T   = \Lam^T \To \Lam + \Lam^T {d \Lam \over d q},  \eqn\threeFiveB
$$
From $\threeFiveA$ ( or by substituting $\threeTwoA$ into the definition of $\kap_1$, i.e, Eq.(B.1)), we have
$$
\kap_1=\left[ \sum_{\U=2}^p(h_\U)^2 \right]^{1/2},    \eqn\threeSixA
$$
and $\threeFiveB$ leads to
$$
 \left(T-\Lam^T {d \Lam\over d q}\right)^2=\Lam^T \left( \To \right)^2 \Lam.  \eqn\threeSixB
$$
Taking the traces of the both sides of $\threeSixB$ and using the anti-symmetry property of $T-\Lam^T d\Lam/dq$, we obtain
$$
\sum_{\U=2}^{p-1}(\kap_\U)^2 = \sum_{\scriptstyle \U,\W=2 \atop \scriptstyle \U > \W}^p \left[ \left(T- \Lam^T {d \Lam \over d q}\right)_{\U\W}\right]^2. \eqn\threeSeven
$$

From $\threeSeven$ we see that we can eliminate the gauge potential $T=(T_{\V\W, 1})$ everywhere along $\M1$ by choosing a  ``gauge" $\Lam$, if and only if the ``torsion"  $\tau$ defined by
$$
 \tau:=\left[ \sum_{\U=2}^{p-1} (\kap_\U)^2\right]^{1/2}   \eqn\threeEight
$$ 
vanishes everywhere along $\M1$.

In the case of $\M1\subset\Real^3$, we obtain from $\threeFiveB$
$$
\omega=\tau-{d \theta \over d q},    \eqn\threeNineA
$$
where $\omega:= -T_{2 3, 1}$ and we take $\Lam$ as
$$
\Lam=\pmatrix{\cos\theta & \sin\theta \cr
             -\sin\theta & \cos\theta} \hbox{  for  }
(\vN_2, \vN_3)=(\vn_2, \vn_3) \Lam.
\eqn\threeNineB
$$

Takagi and Tanzawa$^{\TAKAGI}$ considered the case of a closed curve $\M1$ in the {\it real} space $\Real^3$.
Consider a classical motion of a particle, where a point $\vx(q^1)$ representing a particle position in $\Real^3$ moves along a closed smooth curve;
at the initial point we take $\theta(q_0)=0$, i.e., $(\vN_2(q_0), \vN_3(q_0))=(\vn_2(q_0), \vn_3(q_0))$.
When the particle returns back to the initial point, in the parallel transport$^{\TAKAGI, \KUGLER, \WEINBb}$ with $\omega$ taken to be equal to 0, we obtain
$$
\theta(q_0 + l)=\int_0^l \! dq \, \tau(q),   \eqn\threeTen
$$
which means that $(\vN_2(q_0+l), \vN_3(q_0+l))$ does not always coincide with the initial set generally and differs by the rotation angle given by $\threeTen$.
Here it is to be remembered that, in the case of $\M1\subset \Real^3$, $\tau(q)$ is nonnegative by definition and the Frenet--Serret equations are consistent with the right-handed triad $\{ \vB_1, \vn_2, \vn_3\}$;
$\vn_3=\vB_1\times\vn_2$.(See Appendix B.)

The corresponding situation in quantum mechanics is as follows.$^{\TAKAGI, \FUJIB}$
When we consider the Schr\"odinger equation for a particle on $\M1$, where the effective Hamiltonian on $\M1$ is the one obtained through the thin-tube approximation, the elimination of the gauge term appearing in the momentum operator $\twoSixteenA$ (with the expectation value of $L^{\V\W}$ with respect to the wave function of the normal coordinates $\{ q^\U\}$) causes a phase change of the Schr\"odinger wave function on $\M1$.
This change gives rise to a nonvanishing AB-like effect when $\tau(q)$ satisfies $\threeOne$.

Here we have to remember the remark given at the end of Sec.~2.
In order to reconcile the vanishing $R_{a b ,\V \W}$ in $\M1$ case with the expected nonvanishing AB-like effect $\threeTen$, we have to notice that the condition for $\M1$ to be closed is too stringent.
In the next section, we see that it is enough to assume a closed property of the vector $\vB_1$ in a parameter space in order to obtain the geometrically induced phase $\threeOne$.
Through such a consideration, the relation to the monopole  gauge field as well as the Berry phase will be made clear.

\section{Monopole with a unit strength in the parameter space of $\vB_1(q^1)$}

\noindent
With the aim of examining the problem mentioned above, we make the 3-dimen- \break
sional vector $\vB(q):=\vB_1(q^1)$ correspond to a point particle on a unit sphere $\Sph^2_\Gauss$.
This induces the mapping of $\M1$ onto a curve $C_\vB$ on $\Sph^2_\Gauss$.
Such a correspondence $\M1 \longmapsto C_\vB$ is a kind of so-called {\it Gauss' spherical map}.$^{\KOBAYASHI}$
We use the symbol $\Real^3_\Gauss$ to represent the 3-dimensional Euclidean space with the origin common to $\Sph^2_\Gauss$.

We consider a particle which moves through a distance $l$ along a curve  $\M1$ embedded in $\Real^3$ and, correspondingly,  $\vB$ is assumed to return back to the starting point on $\Sph^2_\Gauss$ after following a simple closed loop $C_\vB$.
In accordance with the Gauss--Bonnet theorem$^{\KOBAYASHI}$ explained shortly in Appendix C, the solid angle $\Omega_\vB$ spanned by $C_\vB$ is given by
$$
|\Omega_\vB|\equiv \int_0^l \tau(q) \, dq  \quad (\hbox{  mod  } 2\pi).   \eqn\threeEleven
$$
When the curve is smooth, $\Omega_\vB$ is equal to the following surface integral over the region $S_\vB$ on $\Sph^2_\Gauss$ enclosed by $C_\vB$ (i.e., $\d S_\vB =C_\vB$):
$$
\SolA = \int_\Surf d\vec\sigma \cdot \vB' 
      = \int_\Surf d\vec\sigma \cdot {\rm rot}\vec A \qquad \hbox{ with  } {\rm rot}\vec A = \vec r/|\vec r|.    \eqn\threeTwelveA
$$
Here, the integrand $\vB'$ represents a unit vector corresponding to a point on $S_\vB$, and  $\vec A$ is the vector field expressed as
$$
\vec A(\vec r)={\pm 1 \over r(r\pm z)}\pmatrix{-y \cr x \cr 0},   \eqn\threeTwelveB
$$
where the negative (positive) $z$-axis of the space $\Real^3_\Gauss$ does not pass through $S_\vB$;
$\vec r=(x, y, z)$ is a position vector in $\Real^3_\Gauss$.
This $\vec A(\vec r)$ is the vector potential generated by a monopole with the unit strength situated at the origin of $\Real^3_\Gauss$.$^{\YANG}$
Thus we see that the phase change of the Schr\"odinger wave function, when a particle moves along $\M1$ (with the nonvanishing torsion) over a distance $l$, is intimately connected, through the relations $\threeTen$, $\threeEleven$ and $\threeTwelveA$, with the tangential line integral of the monopole vector potential $\vec A$ along $C_\vB$ drawn by $\vB$ in $\Real^3_\Gauss$.

The present problem concerning the motion of the vector $\vB$ in $\Real^3_\Gauss$  has the same structure as that investigated by Kugler and Shtrikman,$^{\KUGLER}$ who suggested a classical analogue to Berry's phase.$^{\BERRY}$
They considered a 2-dimensional plane $\bP$ with its normal $\vB$;
a particle undergoes a cylindrically symmetric potential on the plane, on which the particle motion is constrained.
Those authors examined the motion of the particle on that plane $\bP$ when an external force changes $\vB$ slowly along a closed path, so that $\vB$ returns back to its initial place;
thus, evidently they are lead to the same problem of the motion of $\vB$ as ours.
Indeed, they gave Eqs.$\threeTen$ and $\threeEleven$, and mentioned that in many cases the angle $\theta(q_0+l)$ coincides with the Hannay angle.$^{\HANNAY, \BERRY}$

\section{ The case of $\M1$ to be a helix}

\noindent
Let us examine a helical tube around a cylinder.
This tube has a straight circular cross section with its radius $d$, and has a central curve $\M1$ which is a helix with a constant gradient of angle $\beta$ ($0 < \beta < {\pi \over 2}$) around a cylinder with radius $\rho$.
A point on the cylinder is represented by 
$$
\vx(\varphi, z)=(\rho \cos\varphi, \rho\sin\varphi, z).  \eqn\fiveOneA
$$
A parameter $q$ is so chosen as to be a length of the helix, and we write
$$
q=\rho\cdot\varphi/\cos\beta=z/\sin\beta.    \eqn\fiveOneB
$$
We are easy to write explicitly the curvature $\kappa(q)$ and the torsion $\tau(q)$ appearing in the Frenet--Serret equations (see (C.1) in Appendix C) for the triad $\{ \vB(q), \vn_2(q), \vn_3(q) \}$;
$$
\pmatrix{\vB(q) \cr \vn_2(q) \cr \vn_3(q)}=\pmatrix{(-\cos\beta \sin\varphi,& \cos\beta\cos\varphi,& \sin\beta) \cr
(-\cos\varphi,& -\sin\varphi,& 0) \cr
(\sin\beta \sin\varphi,& -\sin\beta\cos\varphi,& \cos\beta) \cr
};      \eqn\fiveTwoA
$$
$\kappa$ and $\tau$ are given by 
$$
\kappa(q)=\cos^2\beta/\rho, \qquad \tau(q)=\cos\beta \sin\beta/\rho.   \eqn\fiveTwoB
$$

When a point on the helix $\M1$, e.g., $P_0(q=0)$, moves along $\M1$ to the point $P'_0(q=2\pi \rho/\cos\beta)$, the vector $\vB$ draws a closed circle $\Curv$ on the unit sphere $S^2_\Gauss$.
(See Fig.~2.)

\centerline{$\lower 6ex \hbox to 150pt {\hrulefill}$}
\centerline{$\hbox to 150pt {\hrulefill}$}
\centerline{Fig.~2}
\centerline{$\hbox to 150pt {\hrulefill}$}
\centerline{$\lower -6ex \hbox to 150pt {\hrulefill}$}

The relation $(C.5)$ on the unit sphere $\Sph^2_\Gauss$ obtained from the Gauss--Bonnet theorem$^{\KOBAYASHI}$ is 
$$
\SolA+\int_{P_0\rightarrow P'_0} \tau(q) dq = 2 \pi.   \eqn\fiveThreeA
$$
(The r.~h.~s. is taken to be equal to $2\pi$, because $\Curv$ tends continuously to a small circle when the angle $\beta$ approaches to $\pi/2$ from below.)
Since
$$
\SolA=2\pi(1-\sin\beta),  \eqn\fiveThreeB
$$
$$
\int_{P_0\rightarrow P'_0}\tau(q) dq = \int_0^{2\pi} {\cos\beta \sin\beta \over \rho} {\rho \over \cos\beta} d\varphi = 2\pi \sin\beta,  \eqn\fiveThreeC
$$
we see the relation $\fiveThreeA$ certainly holds.

Here a remark is given on the vector potential $\vA(\vx)$ in $\Real^3$, the integral of which along the helix leads to the integral such as $\fiveThreeC$.
The condition to be satisfied by $\vA(\vx)$ is
$$
\vB(q) \vA(\vx(q))=\tau(q)       \eqn\fiveFourA
$$
with $\vx(q)$ given by $\fiveOneA$ together with $\fiveOneB$.
Meanwhile, the vector potential due to an infinite solenoid along the $z$-axis and with radius $r_0$ is
$$
\vAso(\vx)=\cases{{\displaystyle {\Phi\over 2\pi}(-{y\over r^2}, \ {x\over r^2},\ 0)} & for $r>r_0$, \cr
{\displaystyle {\Phi\over 2\pi r_0^2}(-y, \ x, \ 0) }& for $r_0\geq r\geq0$;}   \eqn\fiveFourB
$$
thus, by taking $\Phi=2\pi \sin\beta$, on the cylinder surface (with $\rho>r_0$) we obtain
$$
\vAso(\vx(q))={\sin\beta\over \rho} \left(-\sin\varphi,\ \cos\varphi,\ 0 \right)   \eqn\fiveFourC
$$
which satisfies $\fiveFourA$.
Therefore, the geometrically induced gauge potential leads to an effect equivalent to that caused by the above vector potential in $\Real^3$, in which the particle path lies; in this sense,  the effect due to the former gauge structure may be called the geometrically induced AB effect.$^{\TAKAGI}$

The kinetic part of the Hamiltonian in the thin-tube approximation $K^*$ is, from $\twoSeventeenA$, written as 
$$
K^*={1\over 2} p_\V \eta^{\V\W}p_\W + K(\tau) + \Del V^*,   \eqn\fiveFiveA
$$
where
$$
K(\tau):={1\over 2} \left( p_1  - \tau(q) L^{23} \right)^2,  \eqn\fiveFiveB
$$
$$
\Del V^*(q)=-{\hbar^2\over 8} \kappa(q)^2.   \eqn\fiveFiveC
$$
When the total wave function $\Psi(q, q^\U)$ for a scalar particle is assumed to be written as a separate form
$$
\Psi(q, q^\U)= \Phi(q)\cdot f(q^\U)     \eqn\fiveSixA
$$
with $f(q^\U)$ to be the eigenfunction of $L^{23}$
$$
L^{23} f(q^\U)=\hbar m f(q^\U), \qquad m\in \Zahl,   \eqn\fiveSixB
$$
we can write $K(\tau)$ effectively as
$$
\Keff(\tau)={1\over 2}\left(p_1-\hbar m \tau(q) \right)^2. \eqn\fiveSixC
$$
The Schr\"odinger equation for $\Phi(q, t)$ is effectively written as
$$\eqalign{
i\hbar {\d \Phi(q, t) \over \d t} &= \Heff \Phi(q, t), \cr
\Heff &= \Keff(\tau) +\Del V^* + h,
}    \eqn\fiveSixD
$$
where $h$ represents a term determined in accordance with the confining poten- \break
tial.$^{\JENSEN}$

We can eliminate the {\it gauge term} $\hbar m \tau$ by changing the phase of the wave function $\Phi(q)$ as 
$$
\Phi(q)=\left[ \exp\left( i m \int_{q_0}^q \tau(t) d t \right) \right] \Phi_0(q), \eqn\fiveSevenA
$$
where $\Phi_0(q)$ satisfies the Schr\"odinger equation $\fiveSixD$ without the {\it gauge term}.
In the helix case, by taking $q_0=0$ and $q=2\pi \rho/\cos\beta$ we obtain the phase factor in $\fiveSevenA$ as
$$
\Del \Theta(P_0\rightarrow P'_0):= m \! \! \! \! \! \int_0^{2\pi \rho /\cos\beta} \! \! \! \! \! dq {\cos\beta \sin\beta \over \rho} = 2\pi (\sin\beta) \ m,   \eqn\fiveSevenB
$$
which is equal to $\pm \pi$ for $m=\pm 1$ and $\beta=\pi/6$.
$$
\Del \theta(P_0\rightarrow P'_0):=\Del \Theta(P_0\rightarrow P'_0)/m \qquad(\hbox{ when }m\not=0) ,   \eqn\fiveSevenC
$$
is the rotation angle of the frame $\{ \vN_2,\ \vN_3\}$ relative to $\{ \vn_2,\ \vn_3\}$ for the parallel transport from the point $P_0$ to the point $P'_0$.

The mathematical structure of the geometrically induced phase of the helix is similar to that in rotation of photon polarization in a helically wound optical fiber.$^{\CHIAO, \BERRY}$
It is a problem to reconsider the connection of this experiment and our geometrically induced gauge structure, which is left as one of future tasks.

\chapter{Relation of the $\cA$-Connection to the Induced Gauge Field on a Sphere}

\noindent
In this section we investigate, in the case of $\M1\subset \Real^p$ with $p\geq 3$, through the {Gauss} mapping of the vector $\vB_1(q^1)$ the relation of the $\cA$-connection to the induced gauge field on $\Sph^{p-1}$ found by Ohnuki and Kitakado$^{\OHNUKI}$.
Further we note this induced gauge field coincides with the spin-connection on $\Sph^{p-1}$, so that the spin-connection has a monopole-like structure on $\Sph^{p-1}$.

\section{Form of the Wigner rotation in the defining representation}

\noindent
Let us consider the {Gauss} mapping of $\vB_1$, tangent to $\M1\subset\Real^p$, and express it after mapping with the same symbol as $\vB_1(y)$.
Here $y$ is a point on $\Sph^{p-1}_\Gauss$ ($\subset \Real^p_\Gauss$).
Similarly, the normal vectors $ \vN_\U$ ($U=2,\ldots p$) are mapped into the vectors $\vN_\U(y)$ ($U=2,\ldots p$), which construct a basis for the tangent space $T_y \Sph^{p-1}_\Gauss$ at $y$.
The vector $\vB_1(y)$, normal to $T_y \Sph^{p-1}_\Gauss$, is written as
$$
\vB_1(y)=\pmatrix{
y^1/r \cr
y^2/r \cr
\vdots \cr
y^p/r }
\quad \hbox{ with } r=\sqrt{\vy\cdot\vy},   \eqn\quatreOne
$$
and satisfies together with $\{ \vN_\U(y)\ ;\ U=2,\ldots p \}$ the orthonormality and the completeness conditions.
In $\quatreOne$ we take $r\not\equiv 1$ for the later convenience, in other words, every point on $\Sph^{p-1}_\Gauss$ with its radius unity is assumed to be transformed along the radius to a corresponding point on $\Sph^{p-1}_\Gauss$ with its radius  $r\not\equiv 1$.

With the aim of obtaining the induced gauge field on $\Sph^{p-1}$, Ohnuki and Kitakado$^{\OHNUKI}$ examined the Wigner rotation,$^{\WIGNER}$ i.e., the representation of 
$$
\lambda_y:=\alpha_y^{-1} \rho \alpha_{\rho^{-1}y} \quad\hbox{ for }\forall\rho\in SO(p) \hbox{ and } y\in\Real^p.      \eqn\quatreTwoA
$$
Here, $\alpha_y \in SO(p)$ is the ``boost" transformation, satisfying
$$
y=\alpha_y \ \yo \quad\hbox{ with } 
\yo=\pmatrix{
r \cr
0 \cr
\vdots\cr
0
} \in \Real^p.      \eqn\quatreTwoB
$$
The set of $\lambda_y$'s ($y\in \Sph^{p-1}(r)\subset\Real^p$) forms an isotropy group (or the little group$^{\WIGNER}$ according to the Wigner's terminology);
$$
H(\yo):=\left\{ \lambda_y \ \big|\ \lambda_y\in SO(p),\ \lambda_y \yo=\yo \right\}.
\eqn\quatreTwoC
$$
In order to obtain the representation of $\lambda_y$, Ohnuki and Kitakado$^{\OHNUKI}$ set about their consideration by employing the Clifford algebra of order $p$.
For our present purpose, however, it will be seen to be convenient to begin with the defining representation.

We write the matrix $\rho$ for the infinitesimal $SO(p)$-transformation $y\mapsto y'$ = $\rho y$ as
$$
\rho=I+{i\over 2}\sum_{\G, \H=1}^p \omega^{\G \H}S_{\G \H}   \eqn\quatreThreeA
$$
with
$$
\left( S_{\G \H} \right)^\A_{\ \B} =i \left( -\delta_\G^\A \etas{\H \B} + \etas{\G \B} \delta_\H^\A \right).     \eqn\quatreThreeB
$$
Here and in the following, the symbol $\sum_{\G, \H=1}^p$ is omitted for dummy indices.
Any $\alpha_y$ in the defining representation is represented by employing an arbitrary set of the basis vectors $\{ \vN_\U(y)\ ;\ U=2,\ldots , p \}$ of $T_y\Sph^{p-1}_\Gauss$ as 
$$
\alpha_y=\left( (\alpha_y)^\D_{\ \E} \right)=\left( \vB_1(y), \vN_2(y), \cdots , \vN_p(y) \right),     \eqn\quatreFourA
$$
since $\quatreFourA$ satisfies the conditions for $\alpha_y$, i.e., $(i)\ y=\alpha_y \yo$ ($\quatreTwoB$) as well as
$$
(ii)\ \left( \alpha_y \right)^\D_{\ \A} \etas{\D \E} \left( \alpha_y \right)^\E_{\ \B}=\etas{\A \B},    \eqn\quatreFourB
$$
$$
(iii)\ \left( \alpha_y \right)^\D_{\ \A} \eta^{\A \B} \left( \alpha_y \right)^\E_{\ \B}=\eta^{\D \E}.    \eqn\quatreFourC
$$
By using $\quatreThreeA$ and $\quatreFourA$, we can obtain 
$$
\lambda_y=I+{i\over 2} \omega^{\G \H}\pmatrix{
0 & 0 & \cdots & 0 \cr
0 &   &        &   \cr
\vdots& & \vN_\U(y) \cdot  D_{\G \H}(y) \vN_\V(y) & \cr
0 &   &        &   }                          \eqn\quatreFiveA
$$
with 
$$
D_{\G \H}(y):=-i y^\E (\etas{\E \G} \dsub{\H} -\etas{\E \H} \dsub{\G})+  S_{\G \H}.
\eqn\quatreFiveB
$$

\section{The gauge field induced on $\Sph^{p-1}$ and $\cA$-connection}

\noindent
Following Ohnuki and Kitakado,$^{\OHNUKI}$ we write
$$
\lambda_y=I+{i\over 2} \omega^{\G \H} f_{\G \H}(y) =I+i A_\B(y) (\Del y)^\B,
\eqn\quatreSix
$$
where $(\Del y)^\B:={i\over 2}\omega^{\G \H} (S_{\G \H})^\B_{\ \D} y^\D$ = ${1\over 2}\omega^{\G \H} (\delta_\G^\B \etas{\H \E} -\delta_\H^\B\etas{\G \E} ) y^\E$.
This $A_\B$ is the induced gauge field appearing in the combination $p_\B - A_\B$ in the Hamiltonian operator.$^{\OHNUKI}$.
We obtain for $\forall y$ in $\Sph^{p-1}_\Gauss$
$$\eqalign{
A_\B(y) dy^\B =& A_\B(y)\left( \delta_\G^\B -{1\over r^2} y^\B \etas{\G \E} y^\E \right) dy^\G \cr
=& {1 \over r^2} f_{\G \H}(y) y^\H dy^\G \cr
=& {1 \over r^2} dy^\G \pmatrix{
0 & 0 & \cdots & 0 \cr
0 &   &        &   \cr
\vdots& & \vN_\U(y)\cdot D_{\G \H}(y) \vN_\V(y) y^\H & \cr
0 &   &        &   } \cr
=& \pmatrix{
0 & 0 & \cdots & 0 \cr
0 &   &        &   \cr
\vdots& & i \vN_\U(y)\cdot \d_{\B} \vN_\V(y) & \cr
0 &   &        &   }
\left( \delta_\G^\B -{1\over r^2} y^\B \etas{\G \E} y^\E \right)dy^\G,
} \eqn\quatreSevenA
$$
from which we have effectively
$$
\left( A_\B(y) \right)_{\U \V}dy^\B = i \vN_\U \cdot \d_\B \vN_\V(y) dy^\B.  \quad (U, V=2, \ldots p)
\eqn\quatreSevenB
$$
By using a set of curvilinear coordinates $\{ u^j \ ;\ j=2, 3, \ldots, p \}$ on $\Sph^{p-1}_\Gauss\subset \Real^p_\Gauss$, we express $y^\A$ as a function of $u^j$'s; $y^\A=y^\A(u)$.
Then, $\quatreSevenB$ leads to
$$
\left( A_\B(y) \right)_{\U \V} dy^\B = i t_{\U \V, j}(u) du^j    \eqn\quatreSevenC
$$
with 
$$
t_{\U \V, j}(u):=\vN_\U(y(u)) \cdot {\d \over \d u^j}\vN_\V(y(u)).  \eqn\quatreSevenD
$$
The $\cA$-connection $T_{\U \V, 1}(q^1)$ of $\M1$ in the original space $\Real^p$ is given as a pull-back of the $H$-connection (the canonical connection of the second kind) $t_{\U \V, j}$ of $\Sph^{p-1}_\Gauss$ by the {Gauss} mapping, i.e.,
$$
t_{\U \V, j}(u)=T_{\U \V, 1}(q^1) {d q^1\over d s}{d s\over d u^j},  \eqn\quatreEightA
$$
where the parameters $q^1$ and $sr$ are taken to be the lengths along $\M1$ and $C_\vB$ (on $\Sph^{p-1}_\Gauss$), respectively.
When $\vB_1$ and $\vN_\U$'s satisfy Eq.~(B.11) and we take $u^1=sr$, we have
$$\eqalign{
\to_{\U \V, j}(u) =& \To_{\U \V, 1}(q^1){1\over \kappa_1 r} \delta_{j, 1} \cr
=& {\kappa_\V\over \kappa_1 r} \left( \etas{\U, \V+1}-\etas{\U+1, \V}\right) \delta_{j, 1}.
}   \eqn\quatreEightB
$$
In the case of $p=3$, we obtain
$$
\to_{3 2, j}(u)={\tau\over \kappa r}\delta_{j, 1} ={1\over \kappa r}\To_{3 2, 1}(q^1) \delta_{j, 1},    \eqn\quatreEightC
$$
leading to the equation given by (C.6).

Next we give an explicit relation of the $\cA$-connection to the monopole-like gauge field given in Ref. ${\OHNUKI}$.
In order to do this, we choose the concrete forms of $\vN_\U$'s as
$$\eqalign{
\left( \vN_\V(y) \right) \rightarrow & \left( \vphantom{n^B_V}\vn_\V(y) \right)=\left( n_\V^\B(y) \right)\cr
=&\left( -\delta_1^\B{y^\V\over r}+\delta^\B_\W(\delta_\V^\W -{y^\W y^\V\over r \rho_+}) \right)\cr
=& \pmatrix{
-y^\V/r \cr
\delta_\V^\U -{y^\U y^\V\over r \rho_+} }, \quad \rho_+=r+x^1.
}    \eqn\quatreNine
$$
By the way, the choice of $\quatreNine$ is seen to correspond to $\alpha_y$ determined by employing the spinor representation $\alpha_y^{(OK)}$ adopted in Ref.~${\OHNUKI}$;
$$
\gamma_\A \ (y')^\A = \gamma_\B \ \left(\alpha_y\right)^\B_{\ \A} \ y^\A =\alpha_y^{(OK)}\gamma_\A \ y^\A \ \alpha_y^{(OK)\dagger}    \eqn\quatreTenA
$$
for
$$
\alpha_y^{(OK)}:={1\over \sqrt{2 r}}\left[ \sqrt{\rho_+}+\sqrt{\rho_-}{y^\U \gamma_\U\over y_\bot} \gamma_1 \right],    \eqn\quatreTenB
$$
where $\gamma_\A \gamma_\B + \gamma_\B \gamma_\A = 2 \etas{\A \B}$, $\rho_\pm=r\pm y^1$, and
$$
y_\bot =\left( y^\U \etas{\U \V} y^\V \right)^{1/2}=\sqrt{\rho_+ \rho_-}.
$$

By using $\quatreNine$, we obtain
$$\eqalign{
&T_{\W \V, \B}(y)=\vN_\W\cdot\d_\B \vN_\V \cr
=&\cases{ 0 &(for $B=1$ )\cr
{-1\over r \rho_+}\left(\etas{\W \U} \etas{\V \X} - \etas{\V \U}\etas{\W \X} \right) y^\X & (for $B=U$)}
} \eqn\quatreElevenA
$$
$$
=\cases{ 0 &(for $B=1$) \cr
      {-i\over r \rho_+}\left( s_{\U \X} \right)_{\W \U} y^\X &(for $B=U$)};
\eqn\quatreElevenB
$$
here, 
$$
\left( s_{\U \X} \right)_{\W \V}=i\left( -\etas{\U \W} \etas{\X \V} + \etas{\U \V}\etas{\X \W} \right).        \eqn\quatreElevenC
$$
$s_{\U \X}$'s satisfy the $SO(p-1)$-algebra;
$$
\left[ \ssub{\U \X},\ \ssub{\W \Z} \right] = i \left(\etas{\U \W} \ssub{\X \Z}+\etas{\X \Z}\ssub{\U \W}-\etas{\U \Z}\ssub{\X \W}-\etas{\X \W} \ssub{\U \Z} \right).  \eqn\quatreElevenD
$$
Thus, from $\quatreSevenB$ the induced gauge field with the general representation of $s_{\U \X}$ is given by
$$
\left( A_\B(y) \right)=\pmatrix{ 0 \cr {1\over r \rho_+}s_{\U \X} y^\X}. \eqn\quatreElevenE
$$
In the context of Sec.~2, the operator of the induced gauge field appears in the kinetic operator in the form
$$
-{1\over 2}T_{\W \V, \B} L^{\W \V}.   \eqn\quatreTwelveA
$$
In the Schr\"odinger equation, when the angular momentum operator $L^{\W \V}$ operates on the wave function belonging to some irreducible representation specified by $J$, $L^{\W \V}$ is replaced by the matrix $\< J, \alpha | L^{\W \V} |J, \beta\>$ $\equiv$ $(l_{\W \V}^{(J)})_{\alpha \beta}$;
thus, the gauge field appearing in the Hamiltonian of the Schr\"odinger equation is written as
$$\eqalign{
\left( (A_\B(y))_{\alpha \beta} \right) =& \left( -{1\over 2}T_{\W \V, \B}(l^{(J) \W \V})_{\alpha \beta} \right) \cr
=& \pmatrix{ 0\cr {1\over r \rho_+}\left( l^{(J)}_{\U \X} \right)_{\alpha \beta} y^\X },
} \eqn\quatreTwelveB
$$
which is equivalent to $\quatreElevenE$.

\section{Spin-connection on $\Sph^{p-1}$ and the induced gauge field}

\noindent
By employing a set of curvilinear coordinates $\{ u^j \ ;\ j=2, \ldots, p \}$ defined in the subsection {\bf 4.2}, we define
$$
F_j^\A(u):={\d y^\A\over \d u^j }, \quad j=2, 3, \ldots, p.   \eqn\quatreThirteen
$$
We have the two set of vectors (in $\Real^p_\Gauss$), $\{ \vF_j(u) \ ;\ j=2, \ldots, p \}$ and $\{ \vN_\U(y(u))$ ; $U=2, \ldots, p \}$, both of which construct two kinds of bases for the tangent space $T_y \Sph^{p-1}_\Gauss$.
With respect to the suffix $j$, $F_j^\A$ behaves as a covariant vector under the general coordinate transformation of $u^k$ and with respect to the suffix $U$, $N_\U^\A$ behaves as a vector under orthogonal transformation on $T_y \Sph^{p-1}_\Gauss$.

These two kinds of vectors are related with each other by using the {\it vielbein}$^{\SuperG, \WEINBb}$ $h^\U_{\ j}$ as
$$
F^\A_{\ j}=h^\U_{\ j} N_\U^\A, \quad N_\U^\A=h_\U^{\ j} F^\A_{\ j}.    \eqn\quatreFourteenA
$$
The induced metric on $\Sph^{p-1}_\Gauss$, $f_{jk}$, is given by
$$
f_{jk}= F^\A_{\ j} \etas{\A \B} F^\B_{\ k} = h^\U_{\ j} \etas{\U \V} h^\V_{\ k}.
\eqn\quatreFourteenB
$$

Let us examine the structure of the quantity $t_{\U \V, j}$, $\quatreSevenD$.
By utilizing the first fundamental equation of surfaces, $\twoFourA$, we have
$$
\d_j F^\B_{\ k}=\Gamf_{jk} F^\B_{\ m} + H_{1 j k} B^\B_1,    \eqn\quatreFifteenA$$
where $\Gamf_{jk}$ is the Christoffel symbol constructed in term of $f_{jk}$.
Then we obtain
$$\eqalign{
t_{\U\V, j}=& \left( h_\U^{\ l} F^\A_{\ l} \right) \etas{\A \B} \d_j \left( h_\V^{\ k} F^\B_{\ k} \right) \cr
=& h_{\U k} \d_k h_\V^{\ k} + h_{\U m} \Gamf_{jk} h_\V^{\ k}.
}  \eqn\quatreFifteenB
$$
In accordance with the vielbein hypothesis,$^{\SuperG, \WEINBb}$ the total covariant derivative of the vielbein vanishes;
$$
0={\cal D}_j h^\U_{\ k}=\d_j h^\U_{\ k} +\omega^\U_{\ \V, j} h^\V_{\ k}-\Gamf_{kj} h^\U_{\ m},    \eqn\quatreSixteenA
$$
where $\omega^\U_{\ \V, j}$ is the spin-connection.$^{\SuperG, \WEINBb}$
$\quatreSixteenA$ leads to
$$
\omega_{\U \V, j}= \hbox{ R.H.S. of $\quatreFifteenB$ }=t_{\U\V, j}.  \eqn\quatreSIxteenB
$$

Thus we have proved the spin-connection on $\Sph^{p-1}_\Gauss$ coincides with the induced gauge field $t_{\U\V, j}$, and has the monopole-like structure.$^{\OGAWA}$

\chapter{Field Theory of Extended Object}

\noindent
The purpose of this section is to extend the formulation given in Sec.~2 to a field theory, which allows a classical solution.
Although we have to treat infinite-dimensional manifolds on the contrary to the case considered in Sec.~2, it is to be pointed out that essentially the same mathematical structures as those described in Sec.~2 appear in the simple field theory examined below.

\section{Model Lagrangian and field-operator expansion}

\noindent
We take the Lagrangian density expressed as 
$$
\Lag(\phi(x), \d_\mu \phi(x))={-1\over 2}\d_\mu \phi^\A(x) \etas{\A\B} \d^\mu \phi^\B(x)-V(\phi(x)),
\eqn\fourOne
$$
where $(x^\mu)$ is a space-time coordinate;
its metric is Minkowskian $\etas{\mu\nu}$ with $(\etas{\mu\nu})$ = diag$(-1,$ $+1$ , $+1,$ $\cdots)$:
the index $A$ of $\phi^\A$ denotes the internal degrees of freedom;
$\etas{\A\B}$ is the metric tensor in the internal space and assumed not to depend on $\phi^\A$, $\d_\mu \phi^\A$ as well as $(x^\mu)$.
The potential $V(\phi(x))$ is chosen so that a classical solution $\phi_0^\A(\vx)$ exists.

The field operator $\phi^\B(\vx, x^0)$ is assumed to be expanded as $^{\CHRIST - \RAJA}$
$$
\phi^\B(\vx, x^0)=\phi_0^\B(\vx, q^\b(x^0))+\sum_{\U=n+1}^\infty \phi_\U^\B(\vx, q^\b(x^0))q^\U(x^0).
\eqn\fourTwo
$$
Here, $\{ q^1, \ldots, q^n\}$ denotes a set of time-dependent parameters representing the collective coordinates of the classical solution, such as the center-of-mass coordinates of the soliton, the orientation of the internal space and so on;
$\phi_0^\B$ is the classical soliton solution, satisfying
$$
-\vnab_x^2\phi_0^\B(\vx, q)+{\d V \over \d \phi_0^\A(\vx, q)}\eta^{\A\B} =0;   \eqn\fourThreeA
$$
$\d\phi_0^\B(\vx, q)/\d q^\b$ satisfies
$$
\left[ -\vnab_x^2\delta^{\A}_{\D}+{\d^2 V \over \d \phi_0^\D(\vx, q) \d \phi_0^\B(\vx, q)} \eta^{\A \B} \right] \d_\b \phi_0^\D(\vx, q)=0,
    \eqn\fourThreeB
$$
or by putting
$$
{\cal G}^\A_\D[\phi_0]:=\left[ -\vnab_x^2\delta^{\A}_{\D}+  \eta^{\A \B} {\d^2 V \over \d \phi_0^\D(\vx, q) \d \phi_0^\B(\vx, q)} \right],
$$
we can say that $\d_b \phi_0^\D$ is a zero mode solution for ${\cal G}^\A_\D[\phi_0]$;
$\phi_\U^\A(\vx, q)$'s are nonzero mode solutions for ${\cal G}^\A_\D[\phi_0]$, i.e.,
$$
{\cal G}^\A_\D[\phi_0]\phi_\U^\D = \left[ -\vnab_x^2\delta^{\A}_{\D}+ \eta^{\A \B} {\d^2 V \over \d \phi_0^\D(\vx, q) \d \phi_0^\B(\vx, q)} \right]  \phi_\U^\D(\vx, q)=\omega_{\U}^2\phi_\U^\A(\vx, q)
    \eqn\fourThreeC
$$
with $\omega_\U^2 > 0$.
Hereafter we write $\{ \vx, q \}$ simply as $\{ x, q \}$.
$\phi_\V^\B$'s are orthogonal to $\d_\b \phi_0^\A$, and are taken to be normalized to unity, that is, 
$$
\int d\vx \d_\b \phi_0^\B(x, q)\etas{\B\D} \phi_\V^\D(x, q)=0, 
\eqn\fourFourA
$$
$$
\int d\vx  \phi_\U^\B(x, q)\etas{\B\D} \phi_\V^\D(x, q)=\etas{\U\V}. 
\eqn\fourFourB
$$

The expression $\fourTwo$ with these two properties is completely analogous to $\twoOne$.
With the aim of making the analogy clearer, we write $\fourTwo$ as 
$$
\phi^{\B x}(x^0)=\phi_0^{\B x}(q(x^0))+ N_\U^{\ \B x}(q(x^0))\, q^\U(x^0),   \eqn\fourFiveA
$$
and $\fourFourA$ and $\fourFourB$ as
$$
B_\b^{\ \B x}(q)\, \etas{\B x, \D y} N_\V^{\ \D y}(q)=0,    \eqn\fourFiveB
$$
$$
N_\U^{\ \B x}(q)\, \etas{\B x, \D y} N_\V^{\ \D y}(q)=\etas{\U \V}, \eqn\fourFiveC
$$
where $\etas{\B x, \D y}=\etas{\B \D} \del(\vx-\vy)$;
$B_\b^{\B x}(q)$=$\d_\b \phi_0^{\B x}(q)$.
Thus we can think a set of the collective coordinates $\{ q^1, \ldots, q^n\}$ to correspond to a point in $\Mn$, which is embedded in infinite-dimensional {\it Euclidean space} $\Real^\infty$, where we consider a kind of general coordinate transformation from $\{ \phi^{\B x} \}$ to a set of the collective and the normal coordinates $\{ q^\b, q^\U; b=1, \ldots, n, U=n+1, \ldots \}$ with a common time-parameter $x^0$.

\section{Basic relations and the metric}

\noindent
Various relations given in Sec.~2 remain to hold if the index $A$ which has been used to represent the vector property in $\Rp$ is changed to $A x$.
Form $\twoFourA$ and $\twoFourB$, we see the fundamental equations for $B_\b^{\ \A x}$ and $N_\V^{\ \A x}$ are expressed as
$$
\d_\a B_\b^{\ \A x}(q)=\Gam^\rd_{\a\b}(q) B_\rd^{\ \A x}(q)+H_{\W\a\b}(q)\, \eta^{\W\U}N_\U^{\ \A x}(q),    \eqn\fourSixA
$$
$$
\d_\a N_\V^{\ \A x}(q)=-B_\b^{\ \A x}(q)H_{\V \a}^{\ \ \ \b}(q)-T_{\V\W, \a}(q)\, \eta^{\W\U} N_\U^{\ \A x}(q).     \eqn\fourSixB
$$

As in Sec.~2, we use $q^\beta$ to denote a member of a coordinate set $\{ q^\b, q^\U; b=1, \ldots, n, U=n+1, \ldots \}$.
Defining
$$
 \tB_\beta^{\ \A x}(q^\del):=\d \phi^{\A x}/\d q^\beta,    \eqn\FourSeven
$$
we have
$$
\tB_\b^{\ \A x}(q^\del)= B_\b^{\ \A x}(q)+\d_\b N_\V^{\ \A x}(q)\, q^\V,    \eqn\fourSevenA
$$
$$
\tB_\U^{\ \A x}(q^\del)= N_\U^{\ \A x}(q).     \eqn\fourSevenB
$$
A {\it Riemannian metric} on $\Real^\infty$ is given by 
$$
d s^2=d \phi^{\B x} \etas{\B x, \D y} d \phi^{\D y}=\tG_{\alpha \beta}(q^\del) d q^\alpha d q^\beta      \eqn\fourEightA
$$
with
$$
\tG_{\alpha \beta}(q^\del):=\tB_\alpha^{\ \A x}(q^\del)\, \etas{\A x, \B y}\, \tB_\beta^{\ \B y}(q^\del),        \eqn\fourEightB
$$
similarly to $\twoThreeA$.
Meanwhile, the metric tensor on $\Mn$ is given, like $\twoThreeB$, by
$$
g_{\a\b}(q)=B_\a^{\ \A x}(q)\, \etas{\A x, \D y}\, B_\b^{\ \D y}(q).
\eqn\fourNine
$$

\section{Canonical quantization and Hamiltonian operator}

\noindent
We follow the quantization procedure described in the subsection {\bf 2.3}.
The momentum operator $P_{\A x}$ conjugate to $\phi^{\A x}$ is expressed as (cf. $\aOne$)
$$
P_{\A x}={\d \Lag \over \d {\dot \phi}^{\A x}}=\< \d_\alpha \phi^{\B y} \eta_{\B y, \A x}, \dot q^\alpha\> = \< \etas{\A x, \B y} \tB^{\B y}_\alpha,\ \dot q^\alpha \>,
\eqn\fourTen
$$
where ${\dot \phi}^{\A x}:=\d \phi^{\A x}/ \d x^0$.
By expressing the Lagrangian density $\fourOne$ in terms of $\{ q^\beta, \dot q^\beta\}$, we obtain the momentum $p_\beta$, conjugate to $q^\beta$, as (cf. $\twoTwelveD$)
$$
p_\beta :={\d \Lag\over \d \dot q^\beta}=\< \tG_{\beta \alpha}, \dot q^\alpha \>.
\eqn\fourElevenA
$$
By employing $\fourTen$ and $\fourElevenA$, we obtain (cf. $\twoTwelveDB$)
$$
P_{\A x}=\< \tB^\del_{\ \A x}, p_\del \>  \hbox{ with } 
\tB^\del_{\ \A x}:=\etas{\A x, \B y}\tB_\beta^{\ \B y} \tG^{\beta \del},
\eqn\fourElevenB
$$
under the assumption that $[ q^\alpha, \dot  q^\beta ]$ is a function of $(q^\del)$ and not of $( \dot q^\del) $ (cf. $\twoTwelveA$).
Similarly to the subsection {\bf 2.3}, we obtain for the fundamental equal-time commutators
$$
[ \phi^{\A x}, P_{\B y} ]=i\hbar \del^{\A x}_{\B y} \equi [ q^\alpha, \dot q^\beta ] = i\hbar \tG^{\alpha \beta}(q^\del) \equi [ q^\alpha, p_\beta ]=i\hbar \del ^\alpha_\beta.    \eqn\fourTwelveA
$$
Under the condition $[ \d_\alpha, \d_\beta]\phi^{\B x}=0$,
$$
[ P_{\A x}, P_{\B y} ]=0 \equi [p_\alpha, p_\beta]=0 \equi \F^{\alpha \beta}(q^\del)=0.    \eqn\fourTwelveB
$$
Here, $\F^{\alpha \beta}$ is defined in the same way as (A.6) in Appendix A.

Next we consider the Hamiltonian, defined by
$$
H[\phi, P]:=\< P_{\A x}, \dot \phi^{\A x} \> -\int d\vx \Lag,  \eqn\fourThirteenA
$$
where $\Lag$ is given by $\fourOne$.
$H[\phi, P]$ is expressed as
$$
H[\phi, P]={1\over 2} P_{\A x} \eta^{\A x, \B y}P_{\B y}+{1\over 2} (\vnab \phi)^{\A x}\etas{\A x, \B y}(\vnab \phi)^{\B y}+\int V[\phi] d\vx.  \eqn\fourThirteenB
$$
The first term in r.h.s. of $\fourThirteenB$, when expressed in terms of $\{ q^\alpha, p_\alpha \}$-variables, has the same form as $\twoFifteenB$ and is reexpressed as
$$\eqalign{
\tK&={1\over 2} \pi_\a \lambda^{\a\b}\pi_\b+{1\over 2}p_\V \eta^{\V\W}p_\W-{\hbar^2\over 2}\tY(q^\del)  \cr
&=\left( {1\over 2} \pi_a \lambda^{a b} \pi_b +{1\over 2}p_\V \eta^{\V\W}p_\W \right)_{\rm Weyl}-{\hbar^2\over 2}\left[ \tY(q^\del) -{1\over 4} \d_\alpha \d_\beta \tG^{\alpha \beta}\right],
}\eqn\fourFourteenA
$$
where `Weyl' means the Weyl-ordered product.$^{\KUGO, \WEYL}$
By taking into consideration that $\tY$ has the form corresponding to $\twoEighteenB$, the last term is expressed as
$$
-{\hbar^2\over 2}\left[ \tY(q^\del) -{1\over 4} \d_\alpha \d_\beta \tG^{\alpha \beta}\right]={\hbar^2\over 8}\left[ \tR +\tGam^\del_{\alpha \gamma}\tG^{\alpha \beta} \tGam^\gamma_{\beta \del} \right].  \eqn\fourFourteenB
$$
(In the present case, $\tR$=0.)

\section{Separation of the momentum operators corresponding to the collective and the normal coordinates}

\noindent
We will examine the correspondence between the expressions in the present formulation and those derived by Gervais and others.$^{\GERVAIS, \RAJA}$

By utilizing the equality
$$
B_\b^{\ \A x}g^{\b\rd}B_\rd^{\ \B y}+N_\V^{\ \A x}\eta^{\V\W}N_\W^{\ \B y}=\eta^{\A x, \B y},     \eqn\fourFifteen
$$
one can separate $P_{\B x}$ into two parts in the following way;
$$
P_{\B x}={P_0}_{\B x}+S_{\B x}    \eqn\foutSixteenA
$$
with
$$
{P_0}_{\B x}:=\< \etas{\B x, \D z} B_\b^{\ \D z} g^{\b\rd} B_\rd^{\ \A y}, P_{\A y} \>,    \eqn\fourSixteenB
$$
$$
{S}_{\B x}:=\< \etas{\B x, \D z} N_\V^{\ \D z} \eta^{\V\W} N_\W^{\ \A y}, P_{\A y} \>.    \eqn\fourSixteenC
$$
Note that from $p_\V=\< \tB_\V^{\ \A x}, P_{\A x}\>=\< N_\V^{\ \A x}, P_{\A x} \>$,
one obtains
$$
{S}_{\B x}=\< \etas{\B x, \A y} N_\V^{\ \A y} \eta^{\V\W}, p_{\W} \>.    
\eqn\fourSeventeenA
$$
By employing $\fourSixB$--$\fourSevenB$ and remembering $\tG_{\b \V}$ = $\tG_{\V \b}$ = $T_{\V\U, \b}\, q^\U$ and $\tG_{\V\U}$ = $\etas{\V\U}$ ( see $\twoFiveA$), one obtains
$$\eqalign{
\dot \phi^{\B x}(x^0)=&\< \tB_\beta^{\ \B x},\ \dot q^\beta\> \cr
=&\< (\del_\b^\rd-H_{\V\b}^{\ \ \rd}q^\V)B_\rd^{\ \B x},\ \dot q^\b \> \cr
&-\< T_{\V\W, \b}\, q^\V \, \eta^{\W\X}N_\X^{\ \B x},\ \dot q^\b\>+\<N_\X^{\ \B x}\eta^{\X\W}\etas{\W\V},\ \dot q^\V \>  \cr
=& \< (\del_\b^\rd-H_{\V\b}^{\ \ \rd}q^\V)B_\rd^{\ \B x},\ \dot q^\b \> + \<N_\X^{\ \B x}\eta^{\X\W}\tG_{\W\beta},\ \dot q^\beta \>;
}  \eqn\fourSeventeenB
$$
$$
\hbox{the last term = } \<N_\V^{\ \B x}\eta^{\V\W}, p_\W \>=\eta^{\B x, \A y} S_{\A y}. \quad\qquad\qquad\qquad\qquad\qquad \eqn\fourSeventeenC
$$
Thus, ${P_0}_{\B x}$ is expressed as
$$\eqalign{
{P_0}_{\B x}=&\etas{\B x, \A y} \dot \phi^{\A y} -S_{\B x} \cr
=& \< \etas{\B x, \A y}(\del_\b^\rd-H_{\V\b}^{\ \ \rd}q^\V)B_\rd^{\ \A y}, \dot q^\b \>, 
}   \eqn\fourSeventeenD
$$
which can be also confirmed directly from $\fourSixteenB$ by utilizing $P_{\A x}$ = $\< \tB_\alpha^{\B y}\, \etas{\B y, \A x}$, $\dot q^\alpha\>$, i.e., $\fourTen$.
We see ${P_0}_{\B x}$ and $S_{\B x}$ have the following properties:
$$
\< N_\V^{\ \B x}, {P_0}_{\B x} \>=0, \quad \< B_\b^{\ \B x}, {P_0}_{\B x} \>=\< B_\b^{\ \B x}, {P}_{\B x} \>,   \eqn\fourEighteenA
$$
$$
\< N_\V^{\ \B x}, S_{\B x} \>= N_\V^{\ \B x} S_{\B x}=p_\V, \quad \< B_\b^{\ \B x}, {S}_{\B x} \>=B_\b^{\ \B x} {S}_{\B x} = 0. \eqn\fourEighteenB
$$
When we define $\chi^{\B x}$ to be the normal-coordinate part of $\phi^{\B x}$,  i.e.,
$$
\phi^{\B x}=\phi_0^{\B x}+\chi^{\B x} \quad \hbox{ with } \chi^{\B x}:=N_\V^{\B x} q^\V,    \eqn\fourNineteenA
$$
we obtain  due to $\fourFiveB$
$$
\d_\b \phi_0^{\B x} \etas{\B x, \D y}\, \chi^{\D y}=0.  \eqn\fourNineteenB
$$
Note that ${P_0}_{\B x}$ and $S_{\B x}$ are different from $\etas{\B x, \A y}\, \dot \phi_0^{\A y}$ and $\etas{\B x, \A y} \dot \chi^{\A y}$.
In fact, 
$$
\etas{\B x, \A y}\, \dot \phi_0^{\A y} = \< \etas{\B x, \A y}B_\b^{\ \A y}, \dot q^\b\>={P_0}_{\B x}+\Del_{\B x}, \eqn\fourTwentyA
$$
$$
\etas{\B x, \A y} \dot \chi^{\A y} = \< \etas{\B x, \A y}\, \d_\b N_\V^{\ \A y}\, q^\V,\ \dot q^\b\>+\< \etas{\B x, \A y}\, N_\V^{\ \A y},\ \dot q^\V\>
=-\Del_{\B x}+S_{\B x},  \eqn\fourTwentyB
$$
where $\Del_{\B x}:= \etas{\B x, \A y}\< H_{\V\b}^{\ \ \rd} q^\V B_\rd^{\ \A y},\  \dot q^\b\>.$

Now we can prove the important relation for the present consideration;
$$
\< \d_\b \phi^{\B x}, S_{\B x}\>=\< \d_\b \chi^{\B x}, S_{\B x} \>={-1\over 2} T_{\W\V, \b}\,  L^{\W\V},  \eqn\fourTwentyoneA
$$
because 
$$\eqalign{
\hbox{l.h.s.}=&\< B_\b^{\B x} + \d_\b N_\V^{\B x} q^\V,\  \etas{\B x, \A y} N_\W^{\ \A y} \eta^{\W\U}p_\U\> \cr
=& 0-\< T_{\V\W, \b}\, q^\V,\ \eta^{\W\U}p_\U\>
=-\< T_{\V\W, \b},\  \<q^\V\eta^{\W\X},\ p_\X\>\> \cr
=&{-1\over 2}\< T_{\V\W, \b},\ q^\V \eta^{\W\X}p_\X-q^\W \eta^{\V\X}p_\X \> \cr
=& \hbox{r.h.s.} \quad ({\scriptstyle Q.E.D.}).}   \eqn\fourTwentyoneB
$$
Therefore, the operator $\pi_\a$, appearing in $\fourFourteenA$, is expressed as
$$
\pi_\b=p_\b-\< \d_\b \phi^{\B x}, S_{\B x} \>.   \eqn\fourTwentytwo
$$
The classical form of this operator is the same as that given by Gervais and others$^{\GERVAIS}$ in Hamiltonian for the path integral formula, which will be considered in the next subsection.

\section{Path integral formula}

\noindent
In order to obtain the action integral, we first examine the classical form of the operator $\< P_{\A x},\ \dot \phi^{\A x}\>$.
In the $c$-number theory, we have due to $\fourSeventeenB$ and $\fourSeventeenD$
$$\eqalign{
&{P_0}_{\A x} \dot \phi^{\A x} = \dot q^\b (\del_\b^\rd-H_{\V\b}^{\ \ \ \rd}q^\V) B_\rd^{\ \B y} \etas{\B y, \A x} (\del_\a^\e-H_{\W\a}^{\ \ \ \e}q^\W) B_\e^{\ \A x}\dot q^\a    \cr
=&\dot q^\b (\del_\b^\rd-H_{\V\b}^{\ \ \ \rd}q^\V) g_{\rd \e}(\del_\a^\e-H_{\W\a}^{\ \ \ \e}q^\W)\dot q^\a   \cr
=&(p_\rd \tG^{\rd \b}+p_\V \tG^{\V \b})\lambda_{\b\a}\, \dot q^\a;
}\eqn\fourTwentythreeA
$$
by substituting as $\tG^{\rd \b}=\lambda^{\rd\b}$ and $\tG^{\V \b}=-\lambda^{\b \e}T_{\X\W, e}\, q^\W \eta^{\X\V}$,
$\fourTwentythreeA$ is written as
$$\eqalign{
{P_0}_{\A x}\,  \dot \phi^{\A x}
&= \left(p_\rd +{1\over 2}T_{\W\X, \rd} L^{\W\X}\right) \, \dot q^\rd=\pi_\rd \, \dot q^\rd \cr
&= \left(p_\rd -(\d_\rd \chi^{\B y}) S_{\B y}\right)\, \dot q^\rd.
} \eqn\fourTwentythreeB
$$
Meanwhile, $S_{\A x} \dot \phi^{\A x}$ is written as 
$$
S_{\A x} \dot \phi^{\A x} = S_{\A x} \dot \phi_0^{\A x} + S_{\A x}\dot \chi^{\A x} = S_{\A x} \dot \chi^{\A x},   \eqn\fourTewentyA
$$
which is reexpressed as
$$\eqalign{
S_{\A x} \dot \phi^{\A x}&=S_{\A x}\left((\d_\b  \chi^{\A x})\,  \dot q^\b+N_\V^{\ \A x}\, \dot q^\V \right) \cr
&={-1\over 2}T_{\W\V, \b} L^{\W\V}\dot q^\b+p_\V \dot q^\V.
} \eqn\fourTwentyB
$$
Thus we obtain as the $c$-number relations
$$
P_{\A x} \dot \phi^{\A x}=\pi_\rd \dot q^\rd + S_{\A x} \dot \chi^{\A x} \eqn\fourTwentyoneA
$$
$$
\hphantom{P_{\A x} \dot \phi^{\A }}=p_\rd \dot q^\rd + p_\V \dot q^\V, \eqn\fourTwentyoneB
$$

One can write the path-integral formula for the transition amplitude by employing a set of the canonical variables $\{ p_b,\ q^b,\ p_\V,\ q^\V \ ; \ b=1, \ldots , n, V=n+1, \ldots \}$ as integration parameters.
The Hamiltonian utilized in the path-integral formula is obtained from the Hamiltonian operator $\fourThirteenB$ with $\fourFourteenA$ by dropping {\it Weyl} in $\fourFourteenA$.$^{\WEYL}$
With the aim of comparing such a path-integral formula  with that given in Ref.~${\GERVAIS}$ where  parameters $p_a$, $q^b$, $S_{\A x}$, $\chi^{\B y}$ are employed as integration parameters, we note the following point.
First, remember $S_{\B x}$ and $\chi^{\B x}$ are expressed as
$$
S_{\B x}=\etas{\B x, \A y} N_\V^{\ \A y} \eta^{\V \W} p_\W =: N^\W_{\ \B x} p_\W,    \eqn\fourTwoSixA
$$
$$
\chi^{\B x}=N_\V^{\ \B x} q^\V.     \eqn\fourTwoSixB
$$
We cannot regard $\{ p_a$, $q^b$, $S_{\A x}$, $\chi^{\B y} \}$ as a coordinate system on the phase space that is specified by the coordinate system $\{ p_a$, $q^b$, $p_\V$, $q^\W \}$, because $B_b^{\A x} S_{\A x}=0$ and $B_b^{\A x}\etas{\A x, \B y} \chi^{\B y}=0$, i.e., $S_{\A x}$'s and $\chi^{\B y}$'s are not independent on the phase space respectively.
(Compare this with the corresponding situation in the particle case.)
Hence, introducing two sets of new variables $u_\b$ and $v^\b$, $b=1,\ldots, n$, we extend the phase space, and write
$$
S_{\B x}= p_\W N^\W_{\ \B x} + u_\b B^\b_{\ \B x},    \eqn\fourTwoSevenA
$$
$$
\chi^{\B x}=N_\V^{\ \B x} q^\V + B_b^{\ \B x} v^b.     \eqn\fourTwoSevenB
$$
Here $B^\b_{\ \A x}$ is defined by
$$
B^\b_{\ \A x}:=g^{b d} \etas{\A x, \B y} B_d^{\ \B y}.  \eqn\fourTwoSevenC
$$
The original phase space is a subset specified by conditions $u_b$=$v^b$=0, ($b=1,\ldots, n$) in the extended phase space.
The relations $\fourTwoSevenA$ and $\fourTwoSevenB$ are thought to lead to a coordinate transformation
$$
( p_a, \ q^b, \ p_\V, \ q^\W, \ u_c, \ v^d ) \mapsto (p_a, \ q^b, \ S_{\A x}, \ \chi^{\B y})            \eqn\fourTwoEight
$$
on the extended phase space;
this transformation is not canonical as explained in Appendix D.
We can see the transformation to be volume-preserving, i.e.,
$$
\prod_b d p_b d q^b \ \prod_\V d p_\V d q^\V \ \prod_d d u_d d v^d
=\
\prod_b d p_b d q^b \ \prod_{\B x} d S_{\B x} d \chi^{\B x} \ 
.  \eqn\fourThreeOuA
$$
By employing (D.3) in Appendix D, we obtain
$$\eqalign{
& \prod_b d p_b d q^b \ \prod_\V d p_\V d q^\V \ \prod_d d u_d d v^d \ \del(u_d)\ \del(v^d)  \cr
=& \prod_b d p_b d q^b \ \prod_{\B x} d S_{\B x} d \chi^{\B x} \ \prod_d \del(B_d^{\ \A y} \ S_{\A y})\ \del(B^d_{\ \D z}\ \chi^{\D z}).}  \eqn\fourThreeOuB
$$

By noting that the $c$-number $\tK$ to be used at present is, due to $\fourFourteenA$,  expressed  as
$$
\tK={1\over 2}\pi_\a \lambda^{\a\b}\pi_\b+{1\over 2} S_{\A x} \eta^{\A x, \B y} S_{\B y} -{\hbar^2\over 2}\left[ \tY(q^\beta) -{1\over 4}\d_\alpha \d_\beta \tG^{\alpha \beta} \right] \eqn\fourTwentytwoA
$$
with $\pi_\a=p_\a-(\d_\rd \chi^{\B y}) S_{\B y}$ and by utilizing the measure relation $\fourThreeOuB$,
the path integral formula, corresponding to that Eq.~(4.11) in Ref.~$\GERVAIS$, is now written as
$$\eqalign{
\<F\mid I\>_J=&\int [dp_\a dq^\a dS^{\A x} d\chi_{\A x}] \prod_\b \del[\d_\b \phi_0^{\B y} S_{\B y}] \del[g^{b d}\d_d \phi_0^{\B y} \etas{\B y, \D z} \chi^{\D z}] \cr
&\times \exp\biggl[ {i\over \hbar}\int dt\Bigl\{ \left(p_\rd -(\d_\rd \chi^{\B y}) S_{\B y} \right)\, \dot q^\rd +S_{\B y}\dot \chi^{\B y}-H[p_\a, q^\a, S_{\A x}, \chi^{\A x} ]\cr
& +J_{\B y} (\phi_0^{\B y}(q^\a)+\chi^{\B y}(q^\a))\Bigr\} \biggr],
} \eqn\fourTwentytwoB
$$
where
$$
H[p_\a, q^\a, S_{\A x}, \chi^{\A x} ]=\tK+{1\over 2}(\vnab \phi)^{\A x} \etas{\A x, \B y} (\vnab \phi)^{\B y}+\int V[\phi] d\vx.   \eqn\fourTwentytwoC
$$
The expression in R.H.S. of ${\fourTwentytwoB}$ includes (i) the argument functions in the delta function having the definite transformation properties and (ii) $P \dot\phi$-term, different from that in Eq.~(4.11) in Ref.~${\GERVAIS}$.

A few remarks on this formula are to be added as follows.
Firstly, $\tK$ contains the {\it quantum potential} term $\Del V$, i.e., $\fourFourteenB$;
$$\eqalign{
\Del V:=& -{\hbar^2 \over 2}\left[ \tY -{1\over 4} \d_\alpha \d_\beta \tG^{\alpha \beta} \right] \cr
=& {\hbar^2 \over 8} \left[ 2\d_\alpha \left( \tG^{\alpha \beta} \tGam_\beta \right) +\tG^{\alpha \beta} \tGam_\alpha \tGam_\beta +\d_\alpha \d_\beta \tG^{\alpha \beta} \right].
} \eqn\fourThirtytwoA
$$
In the case of one spatial dimension, where $q^1$ is the only collective coordinate included in the combination $x-q^1$, and so $(G_{\alpha \beta})$ and its inverse are $q^1$-independent, it is easy to confirm $\fourThirtytwoA$ to coincide with the expression given by Gervai and Jevicki;$^{\GERVAIS}$
$$\eqalign{
\Del V= & {\hbar^2 \over 8} \biggl[ -{3\over A^2} (\psi^x_1)' \etas{x y} (\psi^y_1)' + {2\over A^3} (\psi_1^x)' \etas{x y} (\phi^y)''+{1\over A^4} (\psi^x_1)' \etas{x y} (\phi^y)' \cr
&\qquad -{1\over A^2}\sum_{\U, \V}\left( N_\U^x \etas{x y} (N_\V^x)' \right)^2 \biggr],
}  \eqn\fourThirtytwoB
$$
where $(\psi^x_1)'=\d \psi_1^x /\d x$, $(\phi_0^x)'=\d \phi_0^x / \d x = -\d \phi_0^x / \d q^1$, $\psi_1^x=(\phi_0^x)'/\sqrt{(\phi_0^y)'\etas{y z}(\phi_0^z)'}$, $A=\psi_1^x \etas{x y} (\phi^y)'$.
$g^{11}$ appearing in the delta function in $\fourTwentytwoB$ is equal to $((\phi_0^x)' \etas{x y} (\phi_0^y)')^{-1}$, the inverse of $g_{11}$;
thus the delta function part is written as
$$
\delta(\psi_1^x \ S_x)\cdot\delta(\psi_1^y \etas{y z} \chi^z),  \eqn\fourThirtythreeA
$$
coinciding with the expression in Ref. ${\GERVAIS}$.
By noting $\tG^{11}=\lambda^{11}$, the inverse of $\lambda_{11}$, to be given by
$$
\lambda^{11}=\left( \psi_1^x \etas{x y} (\phi^y)'\right)^{-2},  \eqn\fourThirtythreeB
$$
the first term in r.h.s. of $\fourTwentytwoA$ is written as
$$
\Bigl( p_1 + (\chi^x)' S_x \Bigr)^2 \bigg/ 2\left( \psi_1^y \etas{y z} (\phi^z)'\right)^2      \eqn\foutThirtythreeC
$$
in accordance with Eq.~(4.12) in Ref.~$\GERVAIS$.
It should be noted, however, that the important difference exist in $P_{\A x} \dot \phi^{\A x}$-term;
from $\fourTwentyoneA$ this term should be written as
$$
\Bigl( p_1 + (\chi^x)' S_x \Bigr) \dot q^1 + S_x \dot \chi^x;  \eqn\fourThirtythreeD
$$
while, the {\it gauge term} $\chi' S$ is dropped in Eq.~(4.11) of Ref.~$\GERVAIS$.
The reason why we obtain such a result exists in the point that the separation of $P_{\A x}$ into $P_{0 \A x}$ and $S_{\A x}$ with the properties $\fourEighteenA$ and $\fourEighteenB$ is definitely performed by employing the fundamental equations for the hypersurface $\fourSixA$ and $\fourSixB$.

It is worthwhile to be remarked that, in theories with spatial-transformation invariance, the gauge term $-(\d_d \chi^{\B y}) S_{\B y}$ is independent of the center-of-mass coordinates as the collective ones.

\chapter{Final Remarks and Discussions}

\noindent
As stressed by Maraner,$^{\MARAB}$ the formalism$^{\FUJIB, \MARA}$ for describing the motion of a particle on $\Mn$ embedded in $\Real^p$ gives a general quantum-mechanical framework for describing dynamical systems with the holonomic constraint as well as those which are treated under the Born--Oppenheimer approximation.$^{\BORN}$
Maraner$^{\MARAB}$ gave some examples of molecular systems to show how the induced-gauge field leads to the monopole structure.
In order to derive the geometrically induced gauge, the thin-tube$^{\TAKAGI}$ (or generally, the thin-layer$^{\FUJIB}$) approximation is employed;
this approximation expressed as $\twoSixA$ means the condition of slow change of geometric properties along $\Mn$ in comparison with that along the direction transverse to $\Mn$, and corresponds to the adiabatic condition.$^{\BORN}$
From the viewpoint of the confining approach to d'Alembert's principle$^{\ARNOLD}$ mentioned in Sec.~1, there exists in quantum theory, apart from the quantum potential of the $\hbar^2$-order, the important contribution, that is, the induced gauge structure coming from the {\it compactified} degrees of freedom.

In Sec.~3, we examined in detail the theoretical meaning  of $\threeOne$ or the integral of the $\cA$-connection in the case of $\M1$ for the purpose of reconciling the vanishing of the {\it field strength} $R_{a b, \W \V}$ and the nonvanishing AB-like effect.
It was pointed out that the nonvanishing phase effect due to the integral of the $\cA$-connection (or the torsion) of $\M1 \subset \Real^3$ has a direct relation to the magnetic flux caused by the monopole with the unit strength which is situated at the origin of the parameter space $\Real^3_\Gauss$ defined through the Gauss' spherical map of the vector $\vB_1$, tangent to $\M1$.
Its structure in $\Real^3_\Gauss$ is completely analogous to the Berry phase found in the mechanical model investigated by Kugler and Shtrikman.$^{\KUGLER}$

Takagi and Tanzawa$^{\TAKAGI}$ gave some considerations concerning the experimental proof of the geometrically induced gauge.
With the same view we examined in the subsection {\bf 3.3} the case of a helical tube, and showed that  in the case of a spinless neutral particle there are mathematical structures corresponding to those in the case of the propagation of polarized photon in a helical optical fiber.$^{\CHIAO}$

In Sec.~4, the relation of the $\cA$-connection in the case of $\M1\subset \Real^p$ to the induced gauge field found by Ohnuki and Kitakado$^{\OHNUKI}$ has made clear through the {Gauss} mapping of $\vB_1$, and this gauge field on $\Sph^{p-1}_\Gauss$ has been proved to be gauge-equivalent to the spin-connection;
the latter has the monopole-like structure.
In the process of the {Gauss} mapping, the tangent and the normal quantities are exchanged.
This suggests that, through such an exchange, it may be possible to estimate the relation of the $\cA$-connection to other induced gauge fields derived on the basis of a certain topological property.
This will be investigated elsewhere.

In Sec.~5 we applied the formalism to the soliton field theory.
Owing to utilizing the fundamental equations for the hypersurface, $\fourSixA$ and $\fourSixB$, the geometrical meaning of the gauge term has been made clear, i.e.,
$$
\d_d \chi^{\B x} \ S_{\B x} = -{1\over 2} T_{\W \V, d} L^{\W \V},  \eqn\FiveOne
$$
and we could easily give the explicit forms of each separated part of $P_{\A x}$, i.e., ${P_0}_{\A x}$ and $S_{\A x}$.
The meaning of the {\it gauge term} $\FiveOne$ in the soliton field theory should be investigated further, especially from the viewpoint mentioned in the preceding paragraph in the finite dimensional case.

It is important in real physical situations to consider a spinor particle.
Two of the present authors (N.~M.~C. and K.~F.) have examined the case $\M1 \subset \Real^3$ in the context of supersymmetry.$^{\CHEPILKO}$
The general formulation for a spinor particle is left as one of future tasks.

Finally we give a remark on a simple  generalization of the formulation described in Sec.~2 to the case $\Mn \subset \Mp$, where the metric depends only on coordinates (and not on their time derivatives).
It is easy for us to derive the extended forms of the fundamental equations for $\Mn$ embedded in $\Mp$,$^{\RASHEV}$ which are written as Eqs.~(E.6) and (E.7) in Appendix E.

Some complication arises, however, when we examine the structure of the metric $\tG_{\alpha \beta}(q,$ $\qU$ ), defined by
$$
\tG_{\alpha \beta}(q, \qU):=\tB_\alpha^{\ \A}(q, \qU) \tG_{\A\D}(X(q, \qU))\tB_\beta^{\ \D}(q, \qU)   \eqn\fiveEightA
$$
instead of $\twoThreeA$.
Here, $\tG_{\A\D}(X)$ is the metric tensor on $\Mp$.
Further we have to define $g_{\a \b}(q)$, the metric tensor on $\Mn$, as
$$
g_{\a \b}(q):=B_\a^{\ \A}(q) G_{\A\D}(q)B_\b^{\ \D}(q)   \eqn\fiveEightB
$$
with $G_{\A\D}(q):=\tG_{\A\D}(X(q, 0))=\tG_{\A\D}(x(q)).$
Thus we see that, in order to examine the gauge structure appearing in the effective Hamiltonian in $\Mn$, we have to expand $\tG_{\A\B}(X(q, \qU))$ a power series of the normal coordinate $\qU$ from the outset.
This is the  origin of the complication.
In spite of such a situation, we can see similar gauge structures to exist, which is explained as follows.

For our purpose, it may be helpful to remember the reason why the kinetic energy $\tK$ given by $\twoFourteenA$ reduces to the form expressed as $\twoFifteenB$.
We write the metric $\tG_{\alpha \beta}$, $\twoFiveA$ as a matrix form
$$
\pmatrix{ \tG_{\alpha \beta}} = \pmatrix{ \alpha_0 & \beta_0^T \cr
                                          \beta_0 & I }.     \eqn\fiveNineA
$$
Its inverse matrix $[ \tG^{\alpha \beta} ]$ is given by 
$$
\pmatrix{ \tG^{\alpha \beta}} =
 \pmatrix{ \xi_0^{-1}       & -\xi_0^{-1} \beta_0^T \cr
         -\beta_0 \xi_0^{-1}& I+\beta_0 \xi_0^{-1} \beta_0^T }, \quad \xi_0:= \alpha_0-\beta_0^T \beta_0.
                                               \eqn\fiveNineB
$$
We see $[ \xi_{0 a b} ]$=$[\lambda_{a b} ]$, so that $[\xi_0^{-1}]$=$[\lambda^{a b}]$;
thus, we have
$$\eqalign{
p_\alpha \tG^{\alpha \beta} p_\beta =& p_a \lambda^{a b} p_b -p_a \lambda^{a b} \beta_{0 b \U} \eta^{\U \V} p_\V \cr
&-p_\U \eta^{\U \V} \beta_{0 \V a} \lambda^{a b} p_b +p_\U \left( \eta^{\U \V} + \eta^{\U \W} \beta_{0 \W a} \lambda^{a b} \beta_{0 b \Z} \eta^{\Z \V}  \right) \cr
=& \left( p_a - p_\U \eta^{\U \W} \beta_{0 \W a} \right) \lambda^{a b} \left( p_b - \beta_{0 b \Z} \eta^{\Z \V} p_\V \right) +p_\U \eta^{\U \V} p_\V
}  \eqn\fiveNineC
$$
and
$$
p_\U \eta^{\U \W} \beta_{0 \W a} = -{1\over 2}T_{\U \W, a} L^{\U \W} = - \beta_{0 a \U} \eta^{\U \W} p_\W,  \eqn\fiveNineD
$$
from which it is easy to see $\tK$ in the case of $\Mn \subset \Real^p$ is expressed as $\twoFifteenB$.

In the case of $\Mn\subset \Mp$, by utilizing the fundamental equation (E.7) together with $\fiveEightB$ and (E.5), we obtain
$$\eqalign{
\tG_{b \V}(q^\beta) =& \tG_{\V b}(q^\beta)=q^\X \left\{ (\d_b N_\X^{\ \A}) G_{\A \B} N_\V^{\ \B} + B_b^{\ \A} N_\X^{\ \D} (\d_\D G_{\A \B} ) N_\V^{\ \B}  \right\} \cr
& \qquad \qquad +\Ord [(q^\W)^2 ] \cr
=& q^\X \left\{ T_{\V \X, b} + B_b^{\ \A} N_\X^{\ \D} N_\V^{\ \B} \Gamma_{\A, \B \D} \right\} +\Ord [(q^\W)^2 ], 
}  \eqn\fiveTenA
$$
$$
\tG_{\U \V}(q^\beta)=\etas{\U \V} + q^\X N_\U^{\ \A} N_\X^{\ \D} N_\V^{\ \B} \left( \Gam_{\A, \B\D} +\Gam_{\B, \A \D}\right) +\Ord[(q^\W)^2].   \eqn\fiveTenB
$$

We write $[\tG_{\alpha \beta}]$ in the present case as
$$
\pmatrix{ \tG_{\alpha \beta} } = 
\pmatrix{ \alpha & \beta^T \cr
          \beta &  \gamma }.         \eqn\fiveTenC
$$
Its inverse is
$$
\pmatrix{ \tG^{\alpha \beta}}=
\pmatrix{\xi^{-1} & -\xi^{-1} \beta^T \gamma^{-1} \cr
        -\gamma^{-1}\beta\xi^{-1} & \gamma^{-1} + \gamma^{-1} \beta \xi^{-1} \beta^T \gamma^{-1} }
        \eqn\fiveTenD
$$
with $\xi := \alpha -\beta^T \gamma^{-1} \beta$.
Similarly to $\fiveNineC$, we have
$$\eqalign{
p_\alpha \tG^{\alpha \beta} p_\beta = & \left( p_a - p_\U (\gamma^{-1} \beta)^{\W}_{\ a} \right) (\xi^{-1})^{a b} \left( p_b - (\beta^T \gamma^{-1})_{b}^{\ \V}  p_\V \right) \cr
& \qquad + p_\U (\gamma^{-1})^{\U \V} p_\V.  
}  \eqn\fiveElevenA
$$
By noting
$$
\alpha_{a b} = \tG_{ a b} = \lambda_{a b} + q^\X q^\Y T_{\U \X, a} \eta^{\U \V} T_{\V \Y, b} + \Ord[(q^\W)^2],     \eqn\fiveElevenB
$$
where $\Ord[(q^\W)^2]$-term includes $\Gam^\A_{\ \B \D}$ and $\d_\E \Gam^\A_{\ \B \D}$, we see $(\xi^{-1})^{a b}$ is equal to $\lambda^{a b}$ plus $\Ord[(q^\W)^2]$-term.
The {\it gauge} term $-p_\U  (\gamma^{-1}\beta)^{\U}_{\ a}$ can be rewritten as
$$
{1 \over 2} L^{\U \V} T_{\U \V, a} - C_a +\Ord[(q^\W)^2]     \eqn\fiveElevenC
$$
with
$$
C_a := p_\U \eta^{\U \V} q^\X B_a^{\ \A} N_\X^{\ \D} N_\V^{\ \B} \Gam_{\A, \B\D}.   \eqn\fiveElevenD
$$

New contributions due to the curved-space property of $\Mp$ appear as the $C_a$ term in $\fiveElevenC$ as well as the metric $(\gamma^{-1})^{\U \V}$ in the last term of $\fiveElevenA$.
In the present stage of consideration it seems enough to point out the structure of the kinetic term in the thin-layer approximation as described above.

The gauge structure investigated in the present paper is a kind of the hidden local symmetries.$^{\CREM - \FUJIA}$
We examined in Sec.~5 a field theoretical model to see how the geometrical gauge structure appears.
The important problem is to explore the mechanism how such a gauge field becomes dynamical, that is, how the kinetic part of the gauge field is derived.
Anyhow, if the geometrically induced gauge structure is confirmed to become dynamical, this idea will suggest an interesting view on the unification  and the origin of the force and the matter.

\vskip 1cm
\noindent{\bf Acknowledgements} \hfill

The present authors  would like to express his thanks to Prof. I.~Tsutsui (INS, Tokyo) and Prof. A.~P.~Kobushkin (ITP, Kiev) for suggestion and discussions.
Thanks are also due to Prof. T.~Okazaki (Hakkaido University of Education, Sapporo) and the members of Elementary Particle Group in Hokkaido University for their interests and discussions.

\APPENDIX{A}{A: Remarks on Canonical Commutation Relations}

\noindent
We give some remarks on equivalence of commutation relations among $\{ q^\beta, p_\beta;$ $ \beta=1, \ldots, p\}$ to those among $\{ X^\A, \dot X^\A; A=1, \ldots, p\}$ in the case where Lagrangian is given by $\Lag=\tK-V(X^\A)$ ($\tK$ is given in $\twoEleven$).
Before doing this, we provide three useful formulae.

For $q$-numbers $A$, $B$ and $C$, as a purely algebraic relation we have
$$
[A,\ \<B,\ C\>]=\< [A,\ B],\ C \> + \< [A,\ C],\ B\>.   \eqn\aLemA
$$
Under the assumption $\twoTwelveA$, for any analytic functions $A(q^\delta)$ and $B(q^\delta)$ of $q^\delta$'s, we have
$$
[A(q^\delta),\ \dot q^\alpha]=i\hbar f^{\beta \alpha}\d_\beta A(q^\delta),
\eqn\aLemB
$$
$$
\< A(q^\delta),\ \<B(q^\delta),\ \dot q^\alpha\> \>=\< A(q^\delta)\ B(q^\delta),\ \dot q^\alpha\>.   
\eqn\aLemC
$$

Now, let us return to the subject.
From the commutation relation $\twoTwelveA$, the definition of $P_\A$ $\aOne$  and the formulae $\aLemA$ and $\aLemB$, we obtain
$$
[ X^\A,\ P_\B]=[X^\A(q^\del),\ \< \etas{\B\D}\, \tB_\beta^{\ \D}(q^\del),\ \dot q^\beta\>]=i\hbar \etas{\B\D}\, \tB_\beta^{\ \D}(q^\del)\, \tB_\gamma^{\ \A}(q^\del)\, f^{\gamma \beta}(q^\del).
$$
Then, we can see
$$
f^{\alpha \beta} = \tG^{\alpha \beta}(q^\del) \equi [X^\A,\ P_\beta]=i\hbar \del^\A_\B.    \eqn\aTwo
$$
The first relation in $\aTwo$ is easily seen to be equivalent to $\twoTwelveE$, i.e.,
$$
[ q^\alpha,\ p_\beta]=i\hbar \del^\alpha_\beta,  \eqn\aThree
$$
where $p_\beta$ is defined by $\twoTwelveD$.

Next we examine commutators such as $[P_\A,\ P_\B]$ and $[p_\alpha,\ p_\beta]$.
Using formulae $\aLemA$ and $\aLemC$, we obtain
$$
[p_\mu,\ p_\nu]=[\< \tG_{\mu \alpha},\ \dot q^\alpha\>,\ \<\tG_{\nu \beta},\ \dot q^\beta\>]=i\hbar\< \tG_{\nu \beta},\ \< \tG_{\mu \alpha},\  \F^{\alpha \beta}\> \>,
$$
where
$$
i\hbar \F^{\alpha \beta}:=[\dot q^\alpha,\ \dot q^\beta]+i\hbar \< \left(-\tG^{\alpha \gamma}\tG^{\beta \del}+\tG^{\beta \gamma} \tG^{\alpha \del} \right) \d_\gamma \tG_{\del \rho},\ \dot q^\rho \>.   \eqn\aFour
$$
From Jacobi identity for $[q^\gamma,\ [\dot q^\alpha,\ \dot q^\beta]]$ we can prove that $\F^{\alpha \beta}$ commutes with  $q^\del$ ($\delta=1, \ldots, p$).$^{\TANIM, \MIYAZ}$
Utilizing this, we obtain
$$
[p_\mu,\ p_\nu]=i\hbar \tG_{\mu \alpha} \tG_{\nu \beta} \F^{\alpha \beta}.
\eqn\aFive
$$
Further, by $\twoTwelveDB$ we have
$$\eqalign{
[p_\alpha,\ p_\beta]&=[\< \tB_\alpha^{\ \A},\  P_\A \>,\ \<\tB_\beta^{\ \B},\ P_\B \>]  \cr
&=i\hbar \<\d_\beta \tB_\alpha^{\ \A}-\d_\alpha \tB_\beta^{\ \A},\ P_\A\>+\<\tB_\alpha^{\ \A},\ \<\tB_\beta^{\ \B},\ [P_\A,\ P_\B]\>\>.  
}\eqn\aSix
$$ 
Noting that $[P_\A,\ P_\B]$ does not contain powers of $\dot q^\del$'s higher than the first, we see under the condition $[\d_\beta,\ \d_\alpha]X^\A(q^\del)=0$
$$
[P_\A,\ P_\B]=0 \equi [p_\alpha,\ p_\beta]=0 \equi \F^{\alpha \beta}=0.
\eqn\aSeven
$$

\APPENDIX{B}{B: Frenet--Serret Equations of a Curve $\M1\subset \Rp$}

\noindent
For the purpose of seeing the situation concretely, we firstly give the Frenet--Serret equations in the case of $\M1\subset \Rp$ with $p\geq 3$, extended to the case of $\M1\subset \Rp$ with $p\geq 3$ from that of $\M1\subset \Real^3$.
Every point on the curve $\M1$ is represented by $\vx(q^1)$, where $q^1$ is taken to represent the length along $\M1$;
thus, $\vB_1(q^1)$ is a unit vector, tangential to the curve $\M1$.
Define the first curvature $\kap_1$ and the first unit vector $\vn_2$, normal to $\vB_1(q^1):={d \vx(q^1)\over d q^1}$, to be
$$
\kap_1=\left[ {d \vB_1\over dq^1}\cdot {d \vB_1\over dq^1} \right]^{1/2},       \eqn\bOne
$$
$$
\vn_2:={1\over \kap_1}{ d\vB_1\over d q^1}.   \eqn\bTwo
$$
Due to $\vn_2\cdot\vn_2=1$ and $\vn_2\cdot \vB_1=0$, we can define the second unit vector $\vn_3$, normal to $\vB_1$ as well as $\vn_2$, to be 
$$
\vn_3:={1\over \kap_2}\left( \kap_1 \vB_1 + { d \vn_2\over d q^1} \right),
\eqn\bThree
$$
where $\kap_2:=|\kap_1 \vB_1 +{d \vn_2 \over d q^1}|$, called the second curvature.
$\bThree$ is arranged in the form as $\threeTwoB$;
$$
{d \over d q^1}\vn_2 = -\kap_1 \vB_1 + \kap_2 \vn_3.
\eqn\bFour
$$
Due to $\vn_3\cdot\vn_3=1$ and $\vn_3\cdot \vB_1=\vn_3\cdot \vn_2=0$, we can define the third unit vector $\vn_4$, normal to $\vB_1$, $\vn_2$ as well as $\vn_3$, to be 
$$
\vn_4={1\over \kap_3}\left( \kap_2 \vn_2 + {d \vn_3 \over d q^1 } \right), 
\eqn\bFive
$$
where the third curvature 
$$
\kap_3:=\left|\kap_2 \vn_2 +{d\vn_3 \over d q^1}\right|.  \eqn\bSix
$$
$\bFive$ is rewritten as 
$$
{d \vn_3 \over d q^1}=-\kap_2 \vn_2 +\kap_3 \vn_4.    \eqn\bSeven
$$
Following the same procedure, we can define the $(p-1)$-th unit vector $\vn_p$, normal to  $\vB_1$, $\vn_1$, $\ldots$, as well as $\vn_{p-1}$, to be
$$
\vn_{p}:={1\over \kap_{p-1}}\left( \kap_{p-2} \vn_{p-2} +{d \over d q^1}\vn_{p-1} \right)  \eqn\bEight
$$
with the $(p-1)$-th curvature $\kap_{p-1}:=|\kap_{p-2} \vn_{p-2} + {d\over d q^1} \vn_{p-1}|$;
$\bEight$ is written as 
$$
{d \over d q^1}\vn_{p-1}= -\kap_{p-2} \vn_{p-2}+\kap_{p-1}\vn_p.   \eqn\bNine
$$
From $\vn_p\cdot\vn_p=1$ and $\vn_p\cdot \vB_1=\vn_p\cdot \vn_2=\cdots=\vn_p\cdot\vn_{p-1}=0$, we obtain that $\left( {d \vn_p \over d q^1} \right)\vn_{p-1}+\kap_{p-1}=0$ and $d \vn_p / d q^1$ is orthogonal to $\vB_1$ as well as all $\vn$'s except for $\vn_{p-1}$;
thus we have 
$$
{d \over d q^1} \vn_p=-\kap_{p-1} \vn_{p-1},  \eqn\bTen
$$

The Frenet--Serret equations in the case of $\M1 \subset \Rp$ (with $p\geq3$) are compactly written as
$$
{d \over d q^1}\left( \vB_1 \ \vn_2 \ \cdots \ \vn_p \right)=\left( \vB_1 \ \vn_2 \ \cdots \ \vn_p \right)\pmatrix{ 0 &-\kap_1 & 0 & \cdots & 0 \cr
                               \kap_1 &      &  &       &       \cr
                                  0   &      & \To &    &       \cr
                               \vdots &      &  &       &       \cr
                                  0   &      &  &       &       }
  \eqn\bEleven
$$
with 
$$
\To:=\pmatrix{0 & -\kap_2 &  0  &    &  &   &  \cr
               \kap_2 &  0 & -\kap_3 &  &  &   &  \cr
               0 &  \kap_3 &  0  &    &  &   &  \cr
                 &       &     &\ddots &   &  \cr
                 &       &     &       & 0 & -\kap_{p-1} \cr
                 &       &     &       & \kap_{p-1}& 0 }.
               \eqn\bTwelve
$$
Similarly, we can examine the detailed structure of the fundamental equations $\twoFourA$ and $\twoFourB$ of the manifold $\Mtwo\subset \Rp$.

\APPENDIX{C}{C: Integral of the Torsion along a Path and Gauss--Bonnet Theorem}

\noindent
The Frenet--Serret equations for a curve $\M1$ in $\Real^3$ are written, in accordance with $\bEleven$, as
$$\eqalign{
{d \vB \over d q} &= \kappa \vn_2, \cr
{d \vn_2 \over d q} &= -\kappa \vB + \tau \vn_3, \cr
{d \vn_3 \over d q} &= -\tau \vn_2,
} \eqn\cOne
$$
where we have set $\vB(q):=\vB_1(q^1)$, $\kappa:=\kappa_1$ and $\tau:=\kappa_2$.
Note that to take $\kappa>0$ and $\tau>0$ is consistent with setting the third member of the triad $\{ \vB, \vn_2, \vn_3 \}$ to be $\vn_3(q)=\vB(q)\times\vn_2(q)$.
The parameter $q$ is taken to be the length along the curve $\M1$ in $\Real^3$.
As explained in the subsection {\bf 3.2}, we consider the Gauss' spherical map$^{\KOBAYASHI}$ $\M1\longmapsto C_\vB$.
We rewrite $\cOne$ with the use of the parameter $s$ which is the length along $C_\vB$ on $\Sph^2_\Gauss$.
Since the curve $C_\vB$ is represented by $\val(s):=\vB(q(s))$ and $d\val/d s$ is a unit vector tangent to $C_\vB$, we obtain from $\cOne$, by taking $d q/d s=1/\kappa$, the following set of equations:
$$\eqalign{
{d \vB \over d s} &=  \vn_2 ={d \val \over d s} \cr
{d \vn_2 \over d s} &= -\vB + {\tau\over \kappa} \vn_3={d^2 \val \over d s^2}, \cr
{d \vn_3 \over d s} &= -{\tau\over \kappa} \vn_2.
} \eqn\cTwo
$$
The normal and the geodesic curvatures,$^{\KOBAYASHI}$ written respectively as $\kappa_n$ and $\kappa_g$, are defined by
$$
{d^2 \val \over d s^2}=\kappa_n \vB + \kappa_g \vn_3, \eqn\cThree
$$
since the vector $\vB(q(s))$ is normal to a tangent plane to the unit sphere $\Sph^2_\Gauss$ at a point on $C_\vB$ specified by the parameter $s$, and $\vn_3(q(s))$ lies in the tangent plane.
Thus we see that 
$$
\kappa_n=-1 \hbox{ and } \kappa_g=\tau/\kappa.     \eqn\cFour
$$

In accordance with the Gauss--Bonnet theorem,$^{\KOBAYASHI}$ we have for a differentiable closed curve $C_\vB$
$$
\Omega_\vB+\int_{C_\vB} \kappa_g(s) ds = 2\pi n, \qquad n \in\Zahl \eqn\cFive
$$
where $\Omega_\vB$ is the solid angle of seeing $C_\vB$, corresponding the region on the left side with respect to the direction of integration along $C_\vB$.
The integral of $\kappa_g$ is rewritten as 
$$
\int_{C_\vB}\kappa_g(s)ds=\int_0^l \tau(q) dq.      \eqn\cSix
$$
When $C_\vB$ can be deformed continuously to a small circle on $\Sph^2_\Gauss$, we take $n=1$.
When $C_\vB$ is not differentiable but piecewise differentiable, 2$\pi$ in r.h.s. of $\cFive$ is necessary to be changed into
$$
2\pi n -\sum_i \varepsilon_i,    \eqn\cSeven
$$
where $\varepsilon_i$ is the outer angle between neighbouring two paths on $C_\vB$ at a nondifferentiable point.

\APPENDIX{D}{D: Properties of the Transformation $\fourTwoEight$}

\noindent
The orthonormality conditions $\fourNine$, $\fourFiveB$ and $\fourFiveC$ are rewritten as
$$
B_b^{\ \A x} \ B^d_{\ \A x} =\del_b^d, \quad B_b^{\ \A x} \ N^\V_{\ \A x} =B^b_{\ \A x} \ N_\V^{\ \A x} =0, \quad N^\W_{\ \A x} \ N_\V^{\ \A x} =\del_\V^\W.   \eqn\fourTwoSevenD
$$
By utilizing $\fourTwoSevenD$, we can solve  $\fourTwoSevenA$ and $\fourTwoSevenB$ inversely, and obtain
$$
p_\W = N_\W^{\ \A x} \ S_{\A x}, \qquad q^\W = N^\W_{\ \A x} \ \chi^{\A x},  \eqn\fourTwoNineA
$$
$$
u_b = B_b^{\ \A x} \ S_{\A x}, \qquad v^b = B^b_{\ \A x} \ \chi^{\A x}.  \eqn\fourTwoNineB
$$
Thus ($p_a$ $q^b$ $S_{\A x}$ $\chi^{\B y}$) is a coordinate system on the extended phase space, and $\fourTwoEight$ is a coordinate transformation on the extended phase space.

The Poisson bracket on the extended phase space is defined by
$$
\{ f,\ g \PB := {\d f \over \d p_\alpha} {\d g \over \d q^\alpha} - {\d f \over \d q^\alpha} {\d g \over \d p_\alpha} + {\d f \over \d u_b} {\d g \over \d v^b} - {\d f \over \d v^b} {\d g \over \d u_b}.          \eqn\dOne
$$
Then
$$
\{ p_\alpha,\ q^\beta \PB =\delta_\alpha^\beta, \qquad \{ u_b,\ v^d \PB =\delta_b^d,        \eqn\dTwo
$$
and Poisson brackets of other combinations of $p_\alpha$, $q^\alpha$, $u_b$ and $v^b$ vanish.
We also have
$$
\{ S_{\A x},\ \chi^{\B y} \PB =\delta_{\A x}^{\B y}, \qquad \{ S_{\A x},\ S_{\B y} \PB = \{ \chi^{\A x},\ \chi^{\B y} \PB=0,    \eqn\dThree
$$
$$
\{ q_a,\ S_{\A x} \PB = 0, \qquad \{ q_a ,\ \chi^{\A x} \PB =0, \eqn\dFour
$$
$$ 
\{ p^a,\ S_{\A x} \PB = p_\W \d_a N^\W_{\ \A x} + u_b \d_a B^b_{\ \A x}, \quad \{ p^a ,\ \chi^{\A x} \PB =q^\W \d_a N_\W^{\ \A x} + v^b \d_a B_b^{\ \A x}. \eqn\dFive
$$
From $\dTwo$ and $\dThree$, $\{ p_a,\ q^a \}$ and $\{ S_{\A x} ,\ \chi^{\B y} \}$ seem to be sets of canonical variables respectively, but because of $\dFive$ the whole set $\{ p_a,\ q^a ,\ S_{\A x} ,\ \chi^{\B y} \}$ is not a set of canonical variables generally.
Therefore the coordinate transformation $\fourTwoEight$ is not canonical generally.

\APPENDIX{E}{E: Fundamental Equations in the Case of $\Mn\subset\Mp$}

\noindent
We can generalize the formulation in the case of $\Mn \subset \Real^p$ $^{\RASHEV}$.
It is needless to mention that the main change arises from the metric tensor $\tG_{\A\B}(X^\D)$ on $\Mp$ appearing instead of $\etas{\A \B}$.

We use the same relations $\twoOne$ and $\twoTwoA$.
The metric tensors in $\Mp$ and $\Mn$ are, instead of $\twoThreeA$ and $\twoThreeB$, now given by
$$
\tG_{\alpha \beta}(q, \qU):=\tB_\alpha^{\ \A}(q, \qU) \tG_{\A\D}(X(q, \qU))\tB_\beta^{\ \D}(q, \qU),   \eqn\eOne
$$
$$
g_{\a \b}(q):=B_\a^{\ \A}(q) G_{\A\D}(q)B_\b^{\ \D}(q),   \eqn\eTwo
$$
where $G_{\A \D}(q)=\tG_{\A \D}(x(q))$.
The quantity $B_\a^{\ \A}(q)$ is assumed to have its rank equal to $n$ when regarded as an $n\times p$ matrix, and behaves as the 1-contravariant (in $\Mp$) and 1-covariant (in $\Mn$) tensor, i.e., under $X^\A \mapsto {X'}^\A={X'}^\A(X)$,
$$
B_\a^\A(q) \mapsto {B'}_\a^{\ \A} (q):={\d {X'}^\A \over \d q^\a}={\d {X'}^\A \over \d X^\B} B_\a^{\ \B}(q);   \eqn\eThree
$$
while, under $q^\b \mapsto {q'}^\b={q'}^\b(q)$ (and $X^\A(q)=X^\A(q(q'))$),
$$
B_\a^{\ \A}(q) \mapsto {B'}_\a^{\ \A}:={\d X^\A \over \d {q'}^\a}={\d q^\b \over \d {q'}^\a} B_\b^{\ \A} (q).      \eqn\eFour
$$
The vectors $\vN_\U(q):=(N_\U^{\ \A}(q))$, $U=n+1, \ldots, p$, normal to $\Mn$ at a point $( x^\A(q) )$ satisfy
$$
B_\a^{\ \A}(q) G_{\A\D}(q) N_\U^{\ \D}(q)=0, \qquad N_\U^{\ \A}(q) G_{\A\D}(q) N_\U^{\ \D}(q)=\etas{\U \V}.       \eqn\eFive
$$

The fundamental equations for $B_\b^{\ \A}$ and $N_\V^{\ \A}$ are expressed as
$$
\nablAst_\a B_\b^{\ \A}(q)=\sum_{\U, \V} H_{\U \a \b}(q) \eta^{\U \V} N_\V^{\ \A}(q),    \eqn\eSix
$$
$$
\nablAst_\a N_\V^{\ \A}(q)=-H_{\U \a}^{\ \ d}(q) B_d^{\ \A}(q) + T_{\W\V, \a}(q) \eta^{\W\U} N_\U^{\ \A}(q),    \eqn\eSeven
$$
where $\nablAst_\a$ represents the absolute derivative with respect to $( q^\a )$ in $\Mn$ (and not with respect to $x^\A$ in $\Mp$);
the defining expressions of these absolute derivatives are written as
$$
\nablAst_\a B_\b^{\ \A}={\d B_\b^{\ \A} \over \d q^\a} + B_\a^{\ \D} \Gam^{\A}_{\ \D \E} B_\b^{\ \E} - \GamabeAst B_\e^{\ \A},    \eqn\eEight
$$
$$
\nablAst_\a N_\U^{\ \A}={\d N_\U^{\ \A} \over \d q^\a} + N_\U^{\ \D} \Gam^{\A}_{\ \D \E} B_a^{\ \E}.    \eqn\eNine
$$

\refout
\figout
\end